\documentclass[twocolumn]{aa}  
\bibpunct{(}{)}{;}{a}{}{,}
\usepackage{txfonts}
\usepackage[version=3]{mhchem}
\begin{document}

\title{\ce{CO2} infrared emission as a diagnostic\\ of planet-forming regions of disks}
\author{Arthur D. Bosman \inst{1} \and Simon Bruderer \inst{2} \and Ewine F. van Dishoeck \inst{1,2} }
\institute{Leiden Observatory, Leiden University, PO Box 9513, 2300 RA Leiden, The Netherlands\\ \email{bosman@strw.leidenuniv.nl}\and
Max-Planck-Insitut f\"{u}r Extraterrestrische Physik, Gie{\ss}enbachstrasse 1, 85748 Garching, Germany}
\date{\today}

\abstract 
{The infrared ro-vibrational emission lines from organic molecules in the inner regions of protoplanetary disks are unique probes of the physical and chemical structure of planet forming regions and the processes that shape them. These observed lines are mostly interpreted with local thermal equilibrium (LTE) slab models at a single temperature. }
{The non-LTE excitation effects of carbon dioxide ({\ce{CO2}}) are studied in a full disk model to evaluate: (i) what the emitting regions of the different {\ce{CO2}} ro-vibrational bands are; (ii) how the {\ce{CO2}} abundance can be best traced using {\ce{CO2}} ro-vibrational lines using future JWST data and; (iii) what the excitation and abundances tell us about the inner disk physics and chemistry. \ce{CO2} is a major ice component and its abundance can potentially test models with migrating icy pebbles across the iceline.}
{A full non-LTE \ce{CO2} excitation model has been built starting from experimental and theoretical molecular data. The characteristics of the model are tested using non-LTE slab models. Subsequently the {\ce{CO2}} line formation has been modelled using a two-dimensional disk model representative of T-Tauri disks where {\ce{CO2}} is detected in the mid-infrared by the {\it Spitzer Space Telescope}.}
{The {\ce{CO2}} gas that emits in the 15 $\mu$m and 4.5 $\mu$m regions of the spectrum is not in LTE and arises in the upper layers of disks, pumped by infrared radiation.
 The $v_2$ 15 $\mu$m feature is dominated by optically thick emission for most of the models that fit the observations and increases linearly with source luminosity. 
 Its narrowness compared with that of other molecules stems from a combination of the low rotational excitation temperature ($\sim 250$ K) and the inherently narrower feature for \ce{CO2}. The inferred \ce{CO2} abundances derived for observed disks range from $3\times 10^{-9}$ to $1\times 10^{-7}$ with respect to total gas density for typical gas/dust ratios of 1000, similar to earlier LTE disk estimates. Line-to-continuum ratios are low, of order a few \%, stressing the need for high signal-to-noise ($S/N > 300$) observations for individual line detections.}
{ The inferred \ce{CO2} abundances are much lower than those found in interstellar ices ($\sim 10^{-5}$), indicating a reset of the chemistry by high temperature reactions in the inner disk. JWST-MIRI with its higher spectral resolving power will allow a much more accurate retrieval of abundances from individual $P-$ and $R-$branch 
lines, together with the \ce{^{13}CO2} $Q$-branch at 15 $\mu$m. The \ce{^{13}CO2} $Q$-branch is particularly sensitive  to possible enhancements of \ce{CO2} due to sublimation of migrating icy pebbles at the iceline(s). Prospects for \textit{JWST}-NIRSpec are discussed as well.}

\keywords{Protoplanetary disks -- molecular processes -- astrochemistry -- radiative transfer -- line: formation}
\authorrunning{A.D. Bosman et al.}
\maketitle 
\section{Introduction}

Most observed exo-planets orbit close to their parent star \citep[for
a review see:][]{Udry2007,Winn2015}.  The atmospheres of these
close-in planets show a large diversity in molecular composition
\citep{Madhusudhan2014} which must be set during planet formation and
thus be representative of the natal protoplanetary
disk. Understanding the chemistry of the inner, planet-forming regions
of circumstellar disks around young stars will thus give us another
important piece of the puzzle of planet formation. Prime molecules for
such studies are \ce{H2O}, CO, \ce{CO2} and \ce{CH4} which are the
major oxygen- and carbon-bearing species that set the overall C/O ratio \citep{Oberg2011}.

The chemistry in the inner disk, i.e., its inner few AU, differs from
that in the outer disk. It lies within the \ce{H2O} and {\ce{CO2}}
icelines so all icy planetesimals are sublimated. The large range of
temperatures (100--1500 K) and densities ($10^{10}-10^{16}$ cm$^{-3}$)
then makes for a diverse chemistry across the inner disk region
\citep[see e.g.][]{Willacy1998,Markwick2002,Agundez2008,Henning2013,Walsh2015}. The driving cause for this
diversity is high temperature chemistry: some molecules such as
\ce{H2O} and HCN have reaction barriers in their formation pathways
that make it difficult to produce the molecule in high abundances at
temperatures below a few hundred Kelvin. As soon as the temperature is high
enough to overcome these barriers, formation is fast and they become
major reservoirs of oxygen and nitrogen. An interesting example is
formed by the main oxygen bearing molecules, \ce{H2O} and {\ce{CO2}}:
the gas phase formation of both these molecules includes the \ce{OH}
radical. At temperatures below $\sim 200$ K the formation of
{\ce{CO2}} is faster, leading to high gas phase abundances, up to
$\sim 10^{-6}$ with respect to (w.r.t.) total gas density, in regions where
{\ce{CO2}} is not frozen out. When the temperature is high enough,
\ce{H2O} formation will push most of the gas phase oxygen into
\ce{H2O} and the \ce{CO2} abundance drops to $\sim10^{-8}$
\citep{Agundez2008,Walsh2014,Walsh2015}. Such chemical transitions can have
strong implications for the atmospheric content of gas giants formed
in these regions if most of their atmosphere is accreted from the
surrounding gas.

A major question is to what extent the inner disk abundances indeed
reflect high temperature chemistry or whether continuously migrating
and sublimating icy planetesimals and pebbles at the icelines
replenish the disk atmospheres \citep{Stevenson1988,Ciesla2006}. Interstellar ices
are known to be rich in \ce{CO2}, with typical abudances of 25\%
w.r.t.\ \ce{H2O} ice, or about $10^{-5}$ w.r.t. total gas density
\citep{deGraauw1996,Gibb2004,Bergin2005,Pontoppidan2008,Boogert2015}.
Cometary ices show similarly high \ce{CO2}/\ce{H2O} abundance ratios
\citep{Mumma2011,LeRoy2015}. Of all molecules with high ice
abundances, \ce{CO2} shows the largest contrast between interstellar
ice and high temperature chemistry abundances, and could therefore be
a good diagnostic of its chemistry.
\cite{Pontoppidan2014} argue based on {\it Spitzer Space Telescope}
data that \ce{CO2} is not inherited from the interstellar medium but
is reset by chemistry in the inner disk. However, that analysis
used a Local Thermodynamic Equilibrium (LTE) \ce{CO2} excitation model coupled with a disk model and did
not investigate the potential of future instruments, which could be
more sensitive to a contribution from sublimating planetesimals. Here
we re-consider the retrieval of \ce{CO2} abundances in the inner
regions of protoplanetary disks using a full non-LTE excitation and
radiative transfer disk model, with a forward look to the new
opportunities offered by the {\it James Webb Space Telescope} (JWST).

The detection of infrared vibrational bands seen from \ce{CO2},
\ce{C2H2} and \ce{HCN}, together with high energy rotational lines of
\ce{OH} and \ce{H2O}, was one of the major discoveries of the
\textit{Spitzer} Space Telescope
\citep[e.g.][]{Lahuis2006,Carr2008,Salyk2008,Salyk2011,Pascucci2009a,Pontoppidan2010,Carr2011,Najita2011,Pascucci2013}. These
data cover wavelengths in the 10--35 $\mu$m range at low spectral resolving power
of $\lambda/\Delta \lambda$=600. Complementary ground-based infrared
spectroscopy of molecules such as CO, OH, \ce{H2O}, \ce{CH4},
\ce{C2H2} and \ce{HCN} also exists at shorter wavelengths in the 3--5
$\mu$m range
\citep[e.g.][]{Najita2003,Gibb2007,Salyk2008,SalykCO2011,Fedele2011,
  Mandell2012,Gibb2013,Brown2013}. The high spectral resolving power of
$R=25000-10^5$ for instruments like Keck/NIRSPEC and VLT/CRIRES have
resolved the line profiles and have revealed interesting kinematical
phenomena, such as disk winds in the inner disk regions
\citep{Pontoppidan2008,Pontoppidan2011,Bast2011,Brown2013}. Further advances are expected with VLT/CRIRES+ as well as through modelling of current data with more detailed physical models. 

Protoplanetary disks have a complex physical structure \citep[see][for
a review]{Armitage2011} and putting all physics, from magnetically
induced turbulence to full radiative transfer, into a single model is
not feasible. This means that simplifications have to be made.
During the \textit{Spitzer} era, the models used to explain the
observations were usually LTE
excitation slab models at a single temperature. With 2D physical
models such as RADLITE \citep{Pontoppidan2009} and with full 2D
physical-chemical models such as Dust and Lines
\citep[DALI,][]{Bruderer2012,Bruderer2013} or Protoplanetary Disk
Model \citep[ProDiMo,][]{Woitke2009} it is now possible to fully take
into account the large range of temperatures and densities as well as
the non-local excitation effects. For example, it has been shown that
it is important to include radiative pumping introduced by hot
($500-1500$ K) thermal dust emission of regions just behind the inner
rim. This has been done for \ce{H2O} by \cite{Meijerink2009} who
concluded that to explain the mid-infrared water lines observed with
\textit{Spitzer}, water is located in the inner $\sim$1 AU in a region
where the local gas-to-dust ratio is 1--2 orders of magnitude higher
than the interstellar medium (ISM) value. \cite{Antonelli2015,Antonelli2016} performed a
protoplanetary disk parameter study to see how disk parameters affect
the \ce{H2O} emission. \cite{Mandell2012} compared a LTE disk model
analysis using RADLITE with slab models and concluded that, while
inferred abundance ratios were similar with factors of a few, there
could be orders of magnitude differences in absolute abundances
depending on the assumed emitting area in slab models \citep[see also
discussion in][]{Salyk2011}. \cite{Thi2013} concluded that the CO
infrared emission from disks around Herbig stars was rotationally cool
and vibrationally hot due to a combination of infrared and
ultraviolet (UV) pumping fields (see also
\citealt{Brown2013}). \cite{Bruderer2015} modelled the non-LTE
excitation and emission of \ce{HCN} concluding that the emitting area
for mid-infrared lines can be 10 times larger in disks than the
assumed emitting area in slab models due to infrared pumping. Our
study of \ce{CO2} is along similar lines as that for HCN.

As {\ce{CO2}} cannot be observed through rotational transitions in the
far-infrared and submillimeter, because of the lack of a permanent dipole
moment, it must be observed through its vibrational transitions at
near- and mid-infrared wavelengths. The {\ce{CO2}} in our own
atmosphere makes it impossible to detect these {\ce{CO2}} lines from
astronomical sources from the ground, and even at altitudes of 13 km
with SOFIA. This means that {\ce{CO2}} has to be observed from space.
{\ce{CO2}} has been observed by \textit{Spitzer} in protoplanetary
disks through its $v_2$ $Q$-branch at 15 $\mu$m where many individual
$Q-$band lines combine into a single broad $Q$-branch feature at low
spectral resolution \citep{Lahuis2006,Carr2008}.  These gaseous CO$_2$
lines have first been detected in high mass protostars and shocks with
the {\it Infrared Space Observatory}
\citep[ISO, e.g.,][]{Dishoeck1996,Boonman2003CO2,Boonman2003}. {\ce{CO2}}
also has a strong band around 4.3 $\mu$m due to the $v_3$ asymmetric
stretch mode. This mode has high Einstein $A$ coefficients and thus
should thus be easily observable, but has not been seen from \ce{CO2}
gas towards protoplanetary disks or protostars, in contrast with the 
corresponding feature in \ce{CO2} ice \citep{Dishoeck1996}.

The {\ce{CO2}} $v_2$ $Q$-branch profile is slightly narrower than that
of \ce{C2H2} and
\ce{HCN} observed at similar wavelengths. These results suggest that {\ce{CO2}} is absent (or strongly
under-represented) in the inner, hottest regions of the disk.  Full
disk LTE modeling of RNO 90 by \cite{Pontoppidan2014} using RADLITE
showed that the observations of this disk favour a low \ce{CO2}
abundance ($10^{-4}$ w.r.t.\ \ce{H2O}, $\approx 10^{-8}$ w.r.t. total gas density). The slab models by \cite{Salyk2011} indicate smaller
differences between the \ce{CO2} and \ce{H2O} abundances, although
\ce{CO2} is still found to be 2 to 3 orders of magnitude lower in
abundance.

To properly analyse \ce{CO2} emission from disks, a full non-LTE
excitation model of the {\ce{CO2}} ro-vibrational levels has to be
made, using molecular data from experiments and detailed quantum
calculations. This model can then be used to perform a simple slab
model study to see under which conditions non-LTE effects may be
important. These same slab model tests are also used to check the
influences of the assumptions made in setting up the ro-vibrational
excitation model. Such \ce{CO2} models have been developed in the past for
evolved AGB stars \citep[e.g.,]{Cami2000,Gonzalez1999} and shocks \citep[e.g.,][]{Boonman2003}, but not applied to disks.

Our \ce{CO2} excitation model is coupled with a full protoplanetary disk
model computed with DALI to investigate the importance of non-LTE excitation, infrared
pumping and dust opacity on the emission spectra.  In addition, the
effects of varying some key disk parameters such as source luminosity
and gas/dust ratios on line fluxes and line-to-continuum ratios are
investigated.  Finally, \textit{Spitzer} data for a set of T-Tauri
disks are analysed to derive the {\ce{CO2}} abundance structure using
parametrized abundances.

\textit{JWST} will allow a big leap forward in the observing
capabilities at near- and mid-infrared wavelengths, where the inner
planet-forming regions of disks emit most of their lines. The
spectrometers on board \textit{JWST}, NIRSPEC and MIRI
\citep{MIRI2015} with their higher
spectral resolving power ($R\approx 3000$) compared to \textit{Spitzer}
($R = 600$) will not only separate many blended lines
\citep{Pontoppidan2010} but also boost line-to-continuum ratios
allowing detection of individual $P$, $Q$ and $R$-branch lines thus
giving new information on the physics and chemistry of the inner
disk. Here we simulate the emission spectra of {\ce{CO2}} and its
\ce{^{13}CO2} isotopologue from a protoplanetary disk at \textit{JWST}
resolution. We investigate which of these lines are most useful for
abundance determinations at different disk heights and point out the
importance of detecting the \ce{^{13}CO2} feature. We also investigate
which features could signify high \ce{CO2} abundances around the
\ce{CO2} iceline due to sublimating planetesimals.

\section{Modelling {\ce{CO2}} emission}

\begin{figure}
\includegraphics[width=\hsize]{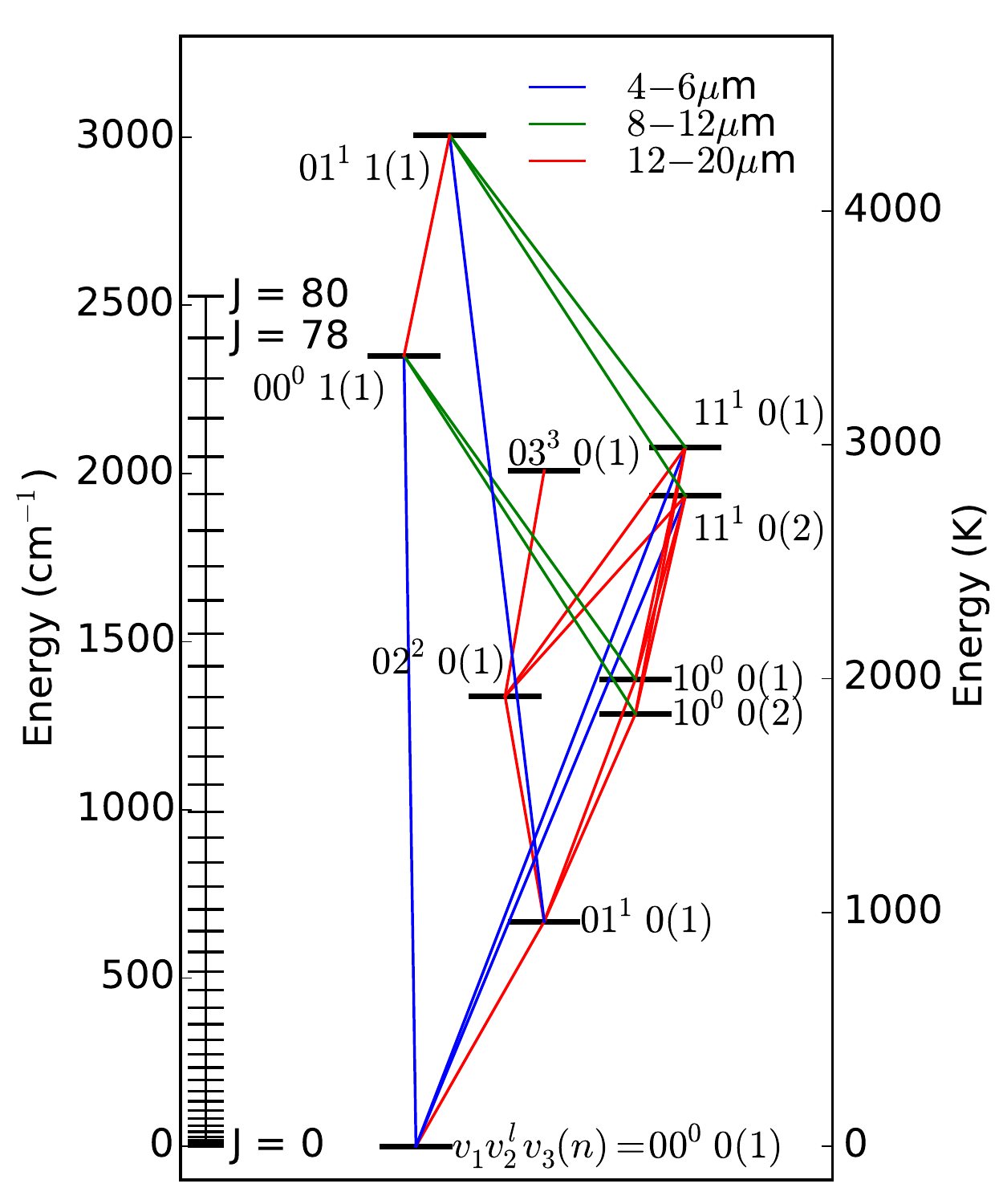}
\caption{\label{fig:energyleveldiagram} Vibrational energy levels of the {\ce{CO2}} molecule (right) together with the rotational ladder of the ground state (left). Note that for the ground state the rotational ladder increases with $\Delta J=2$. Lines connecting the vibrational levels denote the strongest absorption and emission pathways. The colour indicates the wavelength range of the transition: blue, 4--6 $\mu$m, green, 8--12 $\mu$m and red, 12--20 $\mu$m (spectrum in Fig.~\ref{fig:CO2_spectrum}). More information on the rotational ladders is given in Sec.~\ref{ssc:rotlad}.}
\end{figure}

\subsection{Vibrational states}

The structure of a molecular emission spectrum depends on the
vibrational level energies and transitions between these levels that
can be mediated by photons. Fig.~\ref{fig:energyleveldiagram} shows
the vibrational energy level diagram for {\ce{CO2}} from the HITRAN
database \citep{Rothman2013}. Lines denote the transitions that are
dipole allowed. Colours denote the part of the spectrum where features
will show up. This colour coding is the same as in
Fig.~\ref{fig:CO2_spectrum} where a model {\ce{CO2}} spectrum is
presented.

{\ce{CO2}} is a linear molecule with a $^1 \Sigma^{+}_g$ ground
state. It has a symmetric, $v_1$, and an asymmetric, $v_3$, stretching
mode (both of the $\Sigma$ type) and a doubly degenerate bending mode,
$v_2$ ($\Pi$ type) with an angular momentum, $l$. A vibrational state
is denoted by these quantum numbers as: $v_1v_2^lv_3$. The 
vibrational constant of the symmetric stretch mode is very close to
twice that of the bending mode. Due to this
resonance, states with the same value for $2v_1 + v_2$ and the same
angular momentum mix. This mixing leads to multiple vibrational levels
that have different energies in a  process known as Fermi
splitting. The Fermi split levels have the same notation as the
unmixed state with the highest symmetric stretch quantum number, $v_1$
and numbered in order of decreasing energy.\footnote{For example:
  Fermi splitting of the theoretical $02^00$ and $10^00$ levels leads
  to two levels denoted as $10^00(1)$ and $10^00(2)$ where the former
  has the higher energy.}  This leads to the vibrational state
notation of: $v_1v_2^lv_3(n)$ where $n$ is the numbering of the
levels. This full designation is used in
Fig.~\ref{fig:energyleveldiagram}. For the rest of the paper we will
drop the $(n)$ for the levels where there is only one variant.

The number of vibrational states in the HITRAN database is much larger
than the set of states used here. Not all of the vibrational states
are needed to model {\ce{CO2}} in a protoplanetary disk because some
the higher energy levels can hardly be excited, either collisionally or
with radiation, so they should not have an impact on the
emitted line radiation. We adopt the same levels as used for AGB stars
in \cite{Gonzalez1999} and add to this set the $03^30$ vibrational
level.

\subsection{Rotational ladders}
\label{ssc:rotlad}

The rotational ladder of the ground state is given in
Fig.~\ref{fig:energyleveldiagram}. All states up to $J$=80 in each
vibrational state are included; this rotational level corresponds to
an energy of approximately 3700~K (2550~cm$^{-1}$) above the
vibrational state energy. The rotational structure of {\ce{CO2}} is
more complex than that of a linear diatomic like \ce{CO}. This is due
to the fully symmetric wavefunction of {\ce{CO2}} in the ground
electronic state. This means that all states of {\ce{CO2}} need to be
fully symmetric to satisfy Bose-Einstein statistics. As a result, not
all rotational quantum numbers $J$ exist in all of the vibrational
states: some vibrational states miss all odd or all even $J$
levels. There are also additional selections on the Wang parity of the
states ($e$,$f$). For the ground vibrational state this means that
only the rotational states with even $J$ numbers are present and that
the parity of these states is fixed to $e$. 

The rotational structure is summarized in Table~\ref{tab:J_and_parity}. 
The states with $v_2 = v_3 = 0$ all have the same rotational structure as the ground vibrational state.
The $01^10(1)$ state has both even and odd $J$ levels starting
at $J = 1$. The even $J$ levels have $f$ parity, while the odd $J$
levels have $e$ parity. In general for levels with $v_2 \neq 0$ and
$v_3 = 0$, the rotational ladder starts at $J = v_2$ with an even
parity, with the parity alternating in the rotational ladder with
increasing $J$. For $v_3 \neq 0$ and $v_2 = 0$, only odd $J$ levels
exist if $v_3$ is odd, whereas only even $J$ levels exist if $v_3$ is
even. All levels have an $e$ parity. For $v_2 \neq 0$ and $v_3 \neq
0$, the rotational ladder is the same as for the $v_2 \neq 0$ and $v_3
= 0$ case if $v_3$ is even, whereas the parities relative to this case
are switched if $v_3$ is odd. 

\begin{table}
\centering
\caption{\label{tab:J_and_parity} The rotational structure of the vibrational levels included in the model.}
\begin{tabular}{l c l}
\hline \hline
Vibrational level & lowest $J$ & $J$ levels and parity \\
\hline
$00^00(1)$ & 0 & even $J$, $e$\\
$01^10(1)$ & 1 & even $J$, $f$; odd $J$, $e$\\
$02^20(1)$ & 2 & even $J,$ $e$; odd $J$, $f$\\
$10^00(1,2)$ & 0 & even $J$, $e$\\
$03^30(1)$ & 3 & even $J$, $f$; odd $J$, $e$\\
$11^10(1,2)$ & 1 & even $J$, $f$; odd $J$, $e$\\
$00^01(1)$ & 1 & odd $J$, $e$\\ 
$01^11(1)$ & 1 & even $J$, $e$; odd $J$, $f$\\
\hline
\end{tabular}
\end{table}

\subsection{Transitions between states}

To be able to properly model the emission of infrared lines from
protoplanetary disks non-LTE effects need to be taken into
account. The population of each level is determined by the balance of
the transition rates, both radiative and collisional. The radiative
transition rates are set by the Einstein coefficients and the ambient
radiation field. Einstein coefficients for {\ce{CO2}} have been well
studied, both in the laboratory and in detailed quantum chemical
calculations \citep[see e.g.][and references therein]{Rothman2009,Jacquinet2011,Rothman2013,Tashkun2015}.  These are
collected in several databases for {\ce{CO2}} energy levels and
Einstein coefficients such as the Carbon Dioxide Spectroscopic Database
(CDSD) \citep{Tashkun2015} and as part of large molecular databases
such as HITRAN \citep{Rothman2013} and GEISA
\citep{Jacquinet2011}. Here the {\ce{^{12}CO2}} and {\ce{^{13}CO2}} data from
the HITRAN database are used. It should be noted that the differences
between the three databases are small for the lines
considered here, within a few \% in line intensity and less than 1\%
for the line positions.

The HITRAN database gives the energies of the ro-vibrational levels
above the ground state and the Einstein $A$ coefficients of
transitions between them. Only transitions above a certain intensity
at 296 K are included in the databases. The weakest lines included in
the line list are 13 orders of magnitude weaker than the strongest
lines. With expected temperatures in the inner regions of disks
ranging from 100--1000 K, no important lines should be missed
due to this intensity cut. In the final, narrowed down set of states
all transitions that are dipole allowed are accounted for.

Collisional rate coefficients between vibrational states are collected
from literature sources. The measured rate of the relaxation of the
$01^10$ to the $00^00$ state by collisions with \ce{H2} from
\cite{Allen1980} is used. Vibrational relaxation of the $00^01$ state
due to collisions with \ce{H2} is taken from \cite{Nevdakh2003}. For
the transitions between the Fermi split levels the rate by
\cite{Jacobs1975_0200_relax} for collisions between {\ce{CO2}} with
{\ce{CO2}} is used with a scaling for the decreased mean molecular
mass. Although data used here supersede those in \cite{Taylor1969surveyCO2}, that paper does give a sense for the uncertainties of the experiments. The different experiments in \cite{Taylor1969surveyCO2} usually agree within a factor of two and the numbers used here from \cite{Allen1980} and \cite{Nevdakh2003} fall within the spread for their respective transitions. It is thus expected that the accuracy of the individual collisional rate coefficients is better than a factor of two.

No literature information is available for pure rotational transitions
induced by collisions of {\ce{CO2}} with other molecules. We therefore
adopt the \ce{CO} rotational collisional rate coefficients from the
LAMDA database \citep{Schoier2005,Yang2010,Neufeld2012}. Due to the
lack of dipole moment, the critical density for rotational transitions
of {\ce{CO2}} is expected to be very low $(n_\mathrm{crit} < 10^{4})$
cm$^{-3}$ and thus the exact collisional rate
coefficients are not important for the higher density environments
considered here. A method similar to
\cite{Faure2008,Thi2013,Bruderer2015} is used to create the full
state-to-state collisional rate coefficient matrix. The method is
described in Appendix~\ref{app:rovib}.

\subsection{{\ce{CO2}} spectra}

\begin{figure*}
\includegraphics[width=\textwidth]{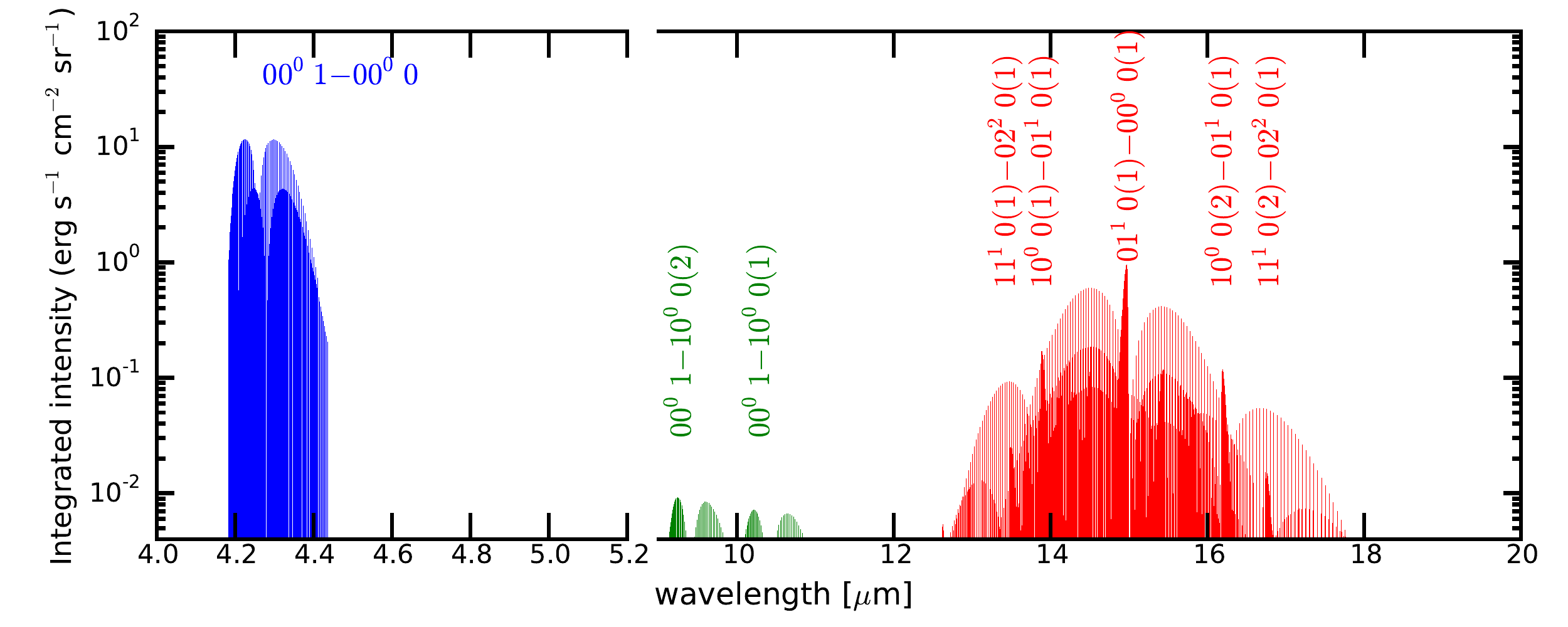}
\caption{\label{fig:CO2_spectrum} {\ce{CO2}} slab model spectrum calculated with RADEX \citep{RADEX}, each line in the spectrum is plotted separately. Slab model parameters are: density, $10^{16}$ cm$^{-3}$; column density of {\ce{CO2}}, $10^{16}$ cm$^{-2}$; kinetic temperature, 750K and linewidth, 1 km s$^{-1}$. For these densities, the level populations are close to local thermal equilibrium (LTE). Spectrum and label color correspond to the colors in Fig.~\ref{fig:energyleveldiagram}}
\end{figure*}

Fig.~\ref{fig:CO2_spectrum} presents a slab model spectrum of {\ce{CO2}} computed using the RADEX program \citep{RADEX}. A density of $10^{16}$ cm$^{-3}$ was used to ensure close
to LTE populations of all levels.  A column density of $10^{16}$
cm$^{-2}$ was adopted, close to the observed value derived by
\cite{Salyk2011}, with a temperature of 750 K and linewidth of 1 km
s$^{-1}$. The transitions are labelled at the approximate location of
their $Q$-branch. The spectrum shows that, due to the Fermi splitting
of the bending and stretching modes, the 15 $\mu$m feature is very
broad stretching from slightly shorter than 12 $\mu$m to slightly
longer than 20 $\mu$m for the absorption in the Earth atmosphere. For astronomical sources, the lines between 14 and 16 $\mu$m are more realistic targets.

Two main emission features are seen in the spectrum. The strong
feature around 4.3 $\mu$m is caused by the radiative decay of the
$00^01$ vibration level to the vibrational ground state. As a
$\Sigma-\Sigma$ transition this feature does not have a $Q-$branch,
but the $R$ and $P$ branches are the brightest features in the
spectrum in LTE at 750 K. The second strong feature is at 15
$\mu$m. This emission is caused by the radiative decay of the $01^10$
vibrational state into the ground state. It also contains small
contributions of the $02^20 \rightarrow 01^10$ and $03^30 \rightarrow
02^20$ transitions. This feature does have a $Q-$branch which has been
observed both in absorption \citep{Lahuis2006} and emission
\citep{Carr2008,Pontoppidan2010}. The \ce{CO2} $Q-$branch is found to
be narrow compared to the other $Q-$branches of \ce{HCN} and \ce{C2H2}
measured in the same sources. 

\begin{figure}
\includegraphics[width = \hsize]{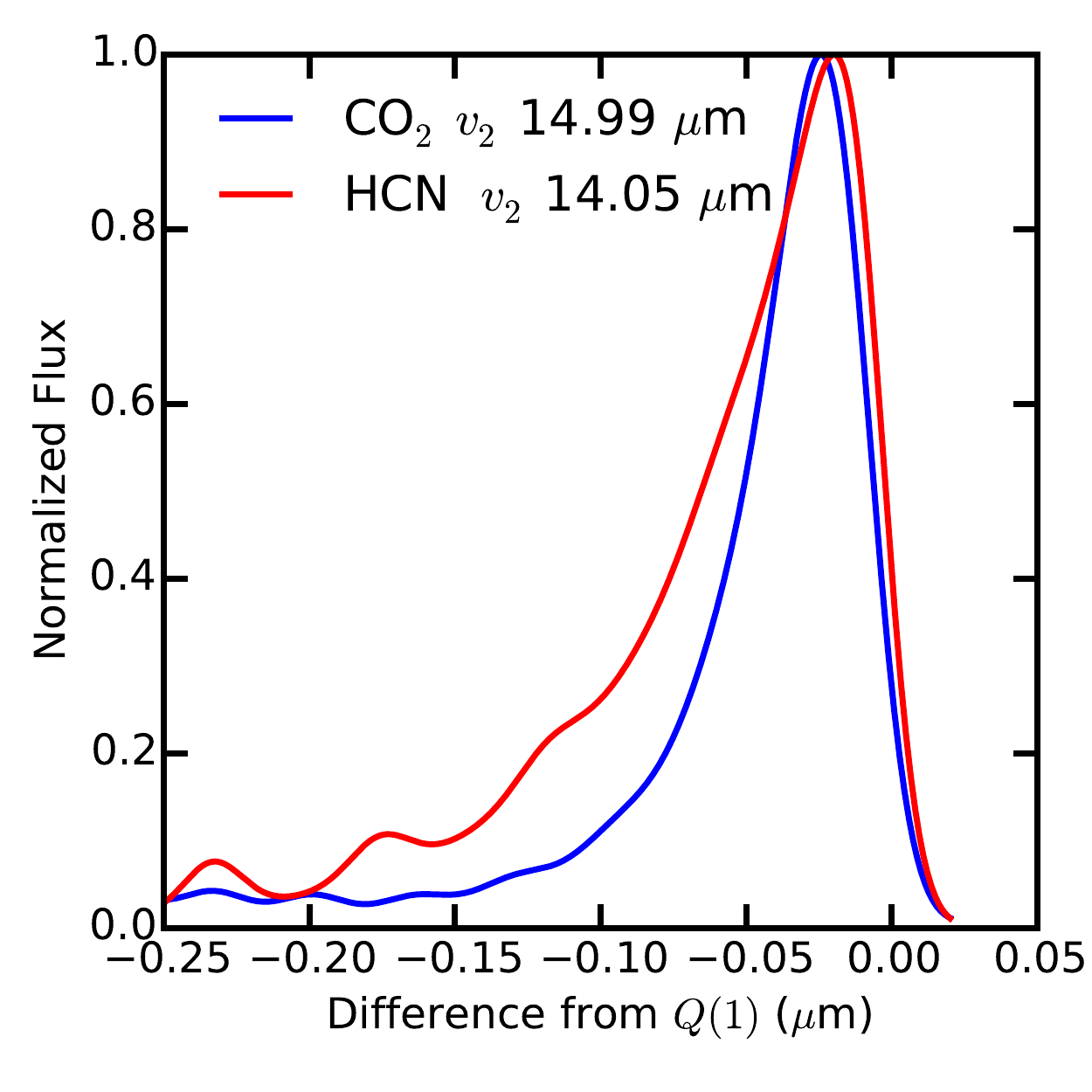}
\caption{\label{fig:CO2_HCN} $v_2$ $Q$-branch profile of {\ce{CO2}} and \ce{HCN} at a temperature of 400 K. Flux is plotted as function of the offset from the lowest energy line (wavelength given in the legend). The lines are convolved to a resolving power $R$=600 appropriate for {\it Spitzer} data. The full width half maximum (FWHM) for {\ce{CO2}} and \ce{HCN} are 0.4 and 0.6 $\mu$m respectively. % in-line with {\ce{CO2}} being heavier and larger.
}
\end{figure}

The narrowness is partly due to the fact that the
{\ce{CO2}} $Q-$branch is intrinsically narrower than the same feature for \ce{HCN}. 
This has to do with the change in the rotational constant 
between the ground and excited vibrational states. A comparison between
$Q-$branch profiles for {\ce{CO2}} and \ce{HCN} for two optically thin
LTE models is presented in Fig.~\ref{fig:CO2_HCN}. The lighter
\ce{HCN} has a full width half maximum (FWHM) that is about 50\%
larger than that of {\ce{CO2}}. The difference in the observed width of the feature is generally larger \citep{Salyk2011}: the \ce{HCN} feature is typically twice as wide as the \ce{CO2} feature. Thus the inferred temperature from the \ce{CO2} $Q-$branch from the observations is low compared to the temperature inferred from the \ce{HCN} feature. The intrinsically narrower \ce{CO2} $Q-$branch amplifies the difference, making it more striking.

\subsection{Dependence on kinetic temperature, density and radiation field}

\begin{figure}
\includegraphics[width = \hsize]{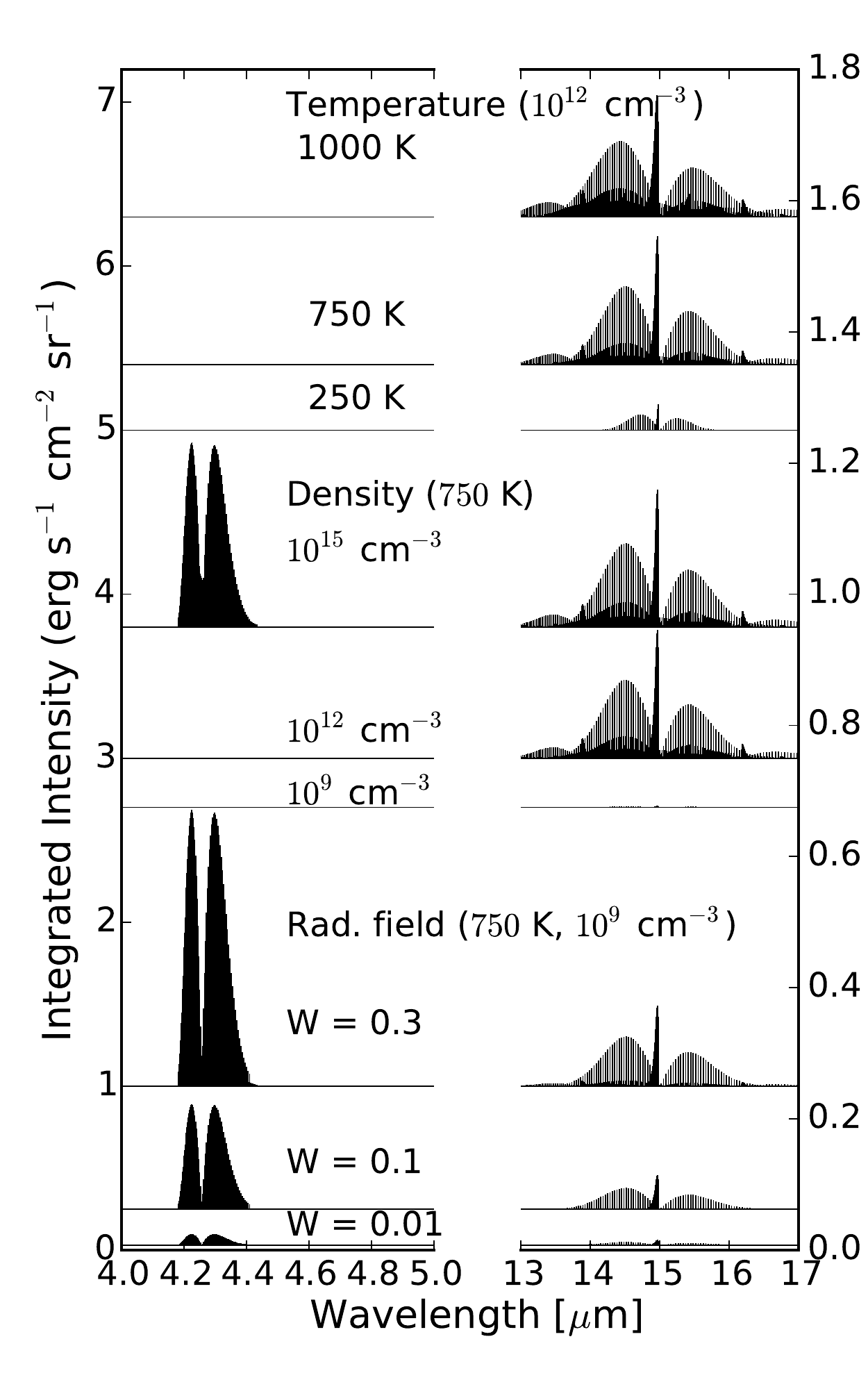}
\caption{\label{fig:CO2_enviroment} {\ce{CO2}} slab model spectra for multiple kinetic temperatures, densities and radiation fields. For all the cases the {\ce{CO2}} column density is kept at $10^{16}$ cm$^{-2}$ and the intrinsic linewidth is set to 1 km s$^{-1}$. The spectra are offset for clarity. All spectra are calculated with RADEX \citep{RADEX}.}
\end{figure}

The excitation of, and the line emission from, a molecule depend
strongly on the environment of the molecule, especially the kinetic
temperature, radiation field and collisional partner density. In
Fig.~\ref{fig:CO2_enviroment} slab model spectra of {\ce{CO2}} for
different physical parameters are compared. The dependence on the
radiation field is modelled by including a blackbody field of 750 K
diluted with a factor $W$: $\langle J_{\nu}\rangle = W
B_\nu(T_{\mathrm{rad}})$ with $T_{\mathrm{rad}} = 750$ K. When testing
the effects of the kinetic temperature and density, no incident
radiation field is included ($W = 0$).

Fig.~\ref{fig:CO2_enviroment} shows that at a constant density of
$10^{12}$ cm$^{-3}$ the 4.3 $\mu$m band is orders of magnitude weaker than the 15 $\mu$m band. The 15 $\mu$m band increases in strength
and also in width, with increasing temperature as higher $J$ levels of
the {\ce{CO2}} $v_2$ vibrational mode can be collisionally
excited. Especially the spectrum at 1000 K shows additional Q branches
from transitions originating from the higher energy $10^00(1)$ and $10^00(2)$ vibrational levels at 14 and 16 $\mu$m.

In the absence of a pumping radiation field, collisions are needed to populate
the higher energy levels. With enough collisions, the excitation temperature becomes 
equal to the kinetic temperature. The density at which the excitation temperature of a
level reaches the kinetic temperature depends on the critical density:
$n_{c} = A_{ul}/K_{ul}$ for a two-level system, where $A_{ul}$ is the
Einstein $A$ coefficient from level $u$ to level $l$ and $K_{ul}$ is
the collisional rate coefficient between these levels. For densities
below the critical density the radiative decay is much faster than the
collisional excitation and de-excitation. This means that the line
intensity scales as $n/n_{c}$. Above the critical density collisional
excitation and de-excitation are fast: the intensity is then no longer
dependent on the density. The critical density of the 15 $\mu$m band
is close to $10^{12}$ cm$^{-3}$, so there is little change in this
band when increasing the density above this value. However, when
decreasing the density below the critical value this results in the
a strong reduction of the band strength. The critical density of the 4.3
$\mu$m feature is close to $10^{15}$ cm$^{-3}$ so below this the lines are orders of magnitude weaker than would be expected from LTE.

Adding a radiation field has a significant impact on both the 4.3 and
15 $\mu$m features. The radiation of a black body of 750 K peaks
around 3.8 $\mu$m so the 4.3 $\mu$m/15 $\mu$m flux ratio in these
cases is larger than the flux ratio without radiation field for
densities below the critical density of the 4.3 $\mu$m lines. Another
difference between the collisionally excited and radiatively excited
states is that in the latter case vibrational levels that cannot be
directly excited from the ground state by photons, such as the
$10^00(1)$ and $10^00(2)$ levels, are barely populated at all.

\section{\ce{CO2} emission from a protoplanetary disk}

To properly probe the chemistry in the inner disk from infrared line
emission one needs to go beyond slab models with their inherent
degeneracies. A protoplanetary disk model such as that used here includes
more realistic geometries and contains a broad range of physical
conditions constrained by observational data. Information
can be gained on the location and extent of the emitting {\ce{CO2}}
region as well as the nature of the excitation process. By comparing
with observational data, molecular abundances can be inferred as
function of location. A critical aspect of the models is the infrared
continuum radiation field, which has to be calculated accurately
throughout the disk. This means that detailed wavelength dependent
dust opacities need to be included and dust temperatures have to be
calculated on a very fine grid, since the pumping radiation can
originate in a different part of the disk than the lines, e.g., the near-infrared for close to the inner rim. The dust is
also important in absorbing some of the line flux, effectively hiding
parts of the disk from our view.

In this section, the {\ce{CO2}} spectra are modelled using the DALI
(Dust and Lines) code \citep{Bruderer2012,Bruderer2013}. The focus is
on emission from the 15 $\mu$m lines that have been observed with
\textit{Spitzer} and will be observable with
\textit{JWST}-MIRI. Trends in the shape of the $v_2$ $Q-$branch and
the ratios of lines in the $P-$ and $R-$branches are investigated and
predictions are presented. First the model and its parameters are
introduced and the results of one particular model are used as
illustration. Finally the effects of various parameters on the
resulting line fluxes are shown, in particular source luminosity and
gas/dust ratio. As in \cite{Bruderer2015}, the model is based on the
source AS 205 (N) but should be representative of a typical T-Tauri
disk. 
\subsection{Model setup}

\label{ssc:setup}
Details of the full DALI model and benchmark tests are reported in \cite{Bruderer2012} and \cite{Bruderer2013}. Here we use the same parts of DALI as in \cite{Bruderer2015}. The model starts with the input of a dust and gas surface density structure. The gas and dust structures are parametrized with a surface density profile 

\begin{equation}
\Sigma(R) = \Sigma_c \left(\frac{R}{R_c}\right)^{-\gamma} \exp\left[-\left(\frac{R}{R_c}\right)^{2-\gamma}\right]
\end{equation}
and vertical distribution
\begin{equation}
\rho(R,\Theta) = \frac{\Sigma(R)}{\sqrt{2\pi}Rh(R)}\exp\left[-\frac12\left(\frac{\pi/2 - \Theta}{h(R)}\right)^2\right],
\end{equation}
with the scale height angle $h(R) = h_c(R/R_c)^\psi$. The values of the parameters for the AS 205 (N) disk are taken from \cite{Andrews2009} who fitted both the SED and submillimeter images simultaneously. As the inferred structure of the disk is strongly dependent on the dust opacities and size distribution, the same values from \cite{Andrews2009} are used. They are summarized in Table~\ref{tab:modelsetup} and the gas density structure is shown in Fig.~\ref{fig:Gridplotoverview}, panel $a$. The central star is a T-Tauri star with excess UV due to accretion. All the accretion luminosity is assumed to be released at the stellar surface as a $10^4$ K blackbody. The density and temperature profile are typical for a strongly flared disk as used here. The temperature, radiation field and \ce{CO2} excitation structure can be found in the appendix, Fig.~\ref{fig:Overview_app}.

In setting up the model special care was taken at the inner rim, where
optical and UV photons are absorbed by the dust over a very short
physical path. To properly get the temperature structure of the disk
directly after the inner rim, high resolution in the radial direction
is needed. Varying the radial width of the first cells showed that the
temperature structure only converges when the cell width of the first
handful of cells is smaller than the mean free path of the UV photons.

The model dust structure is irradiated by the star and the interstellar radiation field. A Monte-Carlo radiative transfer module calculates the dust temperature and the local radiation field at all positions throughout the disk. The gas temperature is then assumed to be equal to the dust temperature. This is not true for the upper and outer parts of the disk. For the regions were \ce{CO2} is abundant in our models the difference between dust temperature and gas temperature computed by self-consistently calculating the chemistry and cooling is less than 5\%. 
The excitation module calculates the {\ce{CO2}} level populations, using a 1+1D escape probablity that includes the continuum radiation due to the dust \citep[Appendix A.2 in][]{Bruderer2013}. Finally the synthetic spectra are derived using the ray tracing module, which solves the radiative transfer equation along rays through the disk. The ray tracing module as presented in \cite{Bruderer2012} is used as well as a newly developed ray-tracing module that is presented in Appendix~\ref{app:fastray} which is orders of magnitude faster, but a few percent less accurate.
In the ray-tracing module a thermal broadening and turbulent broadening with FWHM $\sim0.2$ km s$^{-1}$ is used, which means that thermal broadening dominates above $\sim  40$ K. The gas is in Keplerian rotation around the star. This approach is similar to \cite{Meijerink2009} and \cite{Thi2013} for \ce{H2O} and \ce{CO} respectively. However \cite{Thi2013} used a chemical network to determine the abundances, whereas here only parametric abundance structures are used to avoid the added complexity and uncertainties of the chemical network.

The adopted {\ce{CO2}} abundance is either a constant abundance or a jump abundance profile. The abundance throughout the paper is defined as the fractional abundance w.r.t $n_{\ce{H}} = n(\ce{H}) + 2 n(\ce{H2})$. The inner region is
defined by $T > 200$ K and $A_V > 2$ mag, which is approximately the
region where the transformation of \ce{OH} into \ce{H2O} is faster
than the reaction of \ce{OH} with \ce{CO} to form {\ce{CO2}}. The
outer region is the region of the disk with $T < 200$ K or $A_V < 2$
mag, where the \ce{CO2} abundance is expected to peak.  No \ce{CO2} is
assumed to be present in regions with $A_V < 0.5$ mag as photodissociation is expected to be very efficient in this region.
The physical extent
of these regions is shown in panel $b$ of
Fig.~\ref{fig:Gridplotoverview}.

As shown by \cite{Meijerink2009} and \cite{Bruderer2015}, the
gas-to-dust ("G/D") ratio is very important for the resulting line
fluxes as the dust photosphere can hide a large portion of the
potentially emitting {\ce{CO2}}. Here the gas-to-dust ratio is changed
in two ways, by increasing the amount of gas, or by decreasing the
amount of dust. When the gas mass is increased and thus the dust mass
kept at the standard value of $2.9\times 10^{-4} M_\odot$, this is
denoted by $g/d_{gas}$. If the dust mass is decreased and the gas mass
kept at $0.029 M_\odot$ this is denoted by $g/d_{dust}$.

\begin{table}
\centering
\caption{\label{tab:modelsetup} Adopted standard model parameters for 
the AS 205 (N) star plus disk.}
\begin{tabular}{l l c}
\hline
\hline
Parameter &  & Value\\
\hline
Star\\
Mass & $M_\star$ [$M_\odot$] & 1.0 \\
Luminosity & $L_\star$ [$L_\odot$] & 4.0 \\
Effective temperature & $T_\mathrm{eff}$ [K] & 4250 \\
Accretion luminosity & $L_\mathrm{accr}$ [$L_\odot$] & 3.3 \\
Accretion temperature & $T_\mathrm{accr}$ [K] & 10000 \\
\hline
Disk\\
Disk Mass ($g/d = 100$) & $M_\mathrm{disk}$ [$M_\odot$]& 0.029 \\
Surface density index & $\gamma$ &  0.9 \\
Characteristic radius&  $R_c$ [AU] & 46 \\
Inner radius&  $R_\mathrm{in}$ [AU] & 0.19 \\
Scale height index & $\psi$ & 0.11 \\
Scale height angle & $h_c$ [rad]& 0.18\\
\hline
Dust properties\tablefootmark{a}\\
Size & a[$\mu m$] & 0.005 -- 1000 \\
Size distribution & & $\mathrm{d}n/ \mathrm{d}a \propto a^{-3.5} $\\
Composition & & ISM \\
Gas-to-dust ratio & &100 \\
\hline
Distance & d [pc] & 125 \\
Inclination & $i$ [$^{\circ}$] & 20\\
\hline
\end{tabular}
\tablefoot{
\tablefoottext{a}{Dust properties are the same as those used in \cite{Andrews2009} and \cite{Bruderer2015}. Dust composition is taken from \cite{Draine1984} and \cite{Weingartner2001}.}
}
\end{table}

\subsection{Model results}

\begin{figure*}
	\sidecaption
	\includegraphics[width = 12cm]{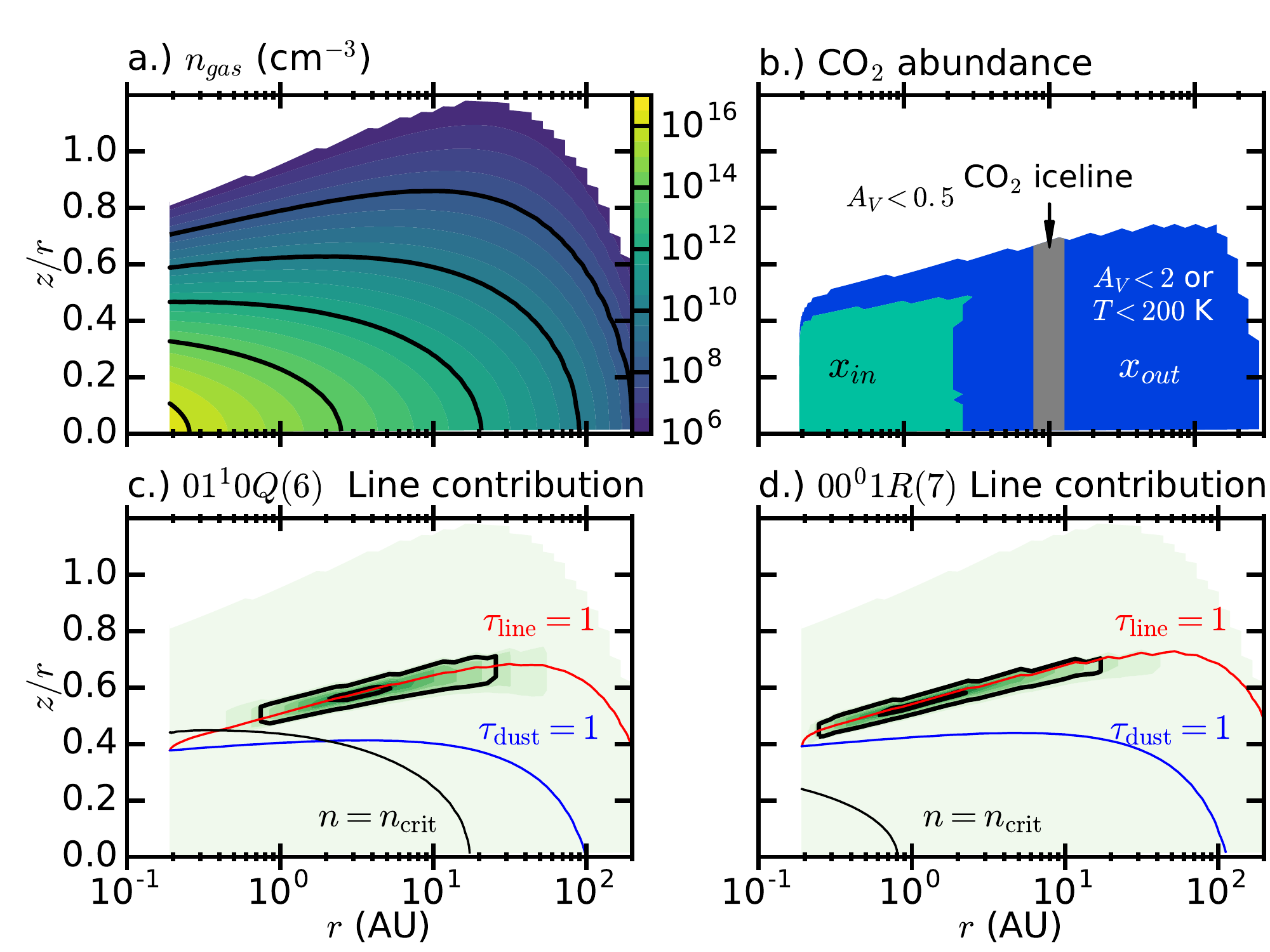}
	\caption{\label{fig:Gridplotoverview} Overview of one of the DALI models showing the disk structure, abundance structure and emitting regions for the Q(6) $01^10$ and R(7) $00^01$ lines. The model shown has a gas-to-dust ratio, $g/d_{\mathrm{gas}} = 1000$ and a constant {\ce{CO2}} abundance of $10^{-7}$ with respect to \ce{H}. The panels show: ($a$) gas density structure; ($b$) abundance structure used models: $x_{\mathrm{in}}$ and $x_{\mathrm{out}}$ are the {\ce{CO2}} abundances in the inner and outer region respectively, the grey region is part of the outer region and denotes the region around the \ce{CO2} iceline where planetesimals are assumed to vaporize. The abundance in this region is varied in the models in Sec.~\ref{ssc:Iceline}; ($c$) line contribution function of the $Q$(6) $01^10$ line at 15 $\mu$m, the contours show the areas in which 25\% and 75\% of the total flux is emitted; ($d$) contribution function for the $R$(7) $00^01$ line at 4.3 $\mu$m. Panels $c$ and $d$ have the $\tau = 1$ surface of dust (blue) and line (red) and the $n=n_{\textrm{crit}}$ surface (black) overplotted for the relevant line.}
\end{figure*}

\begin{figure}
\centering
\includegraphics[width=\hsize]{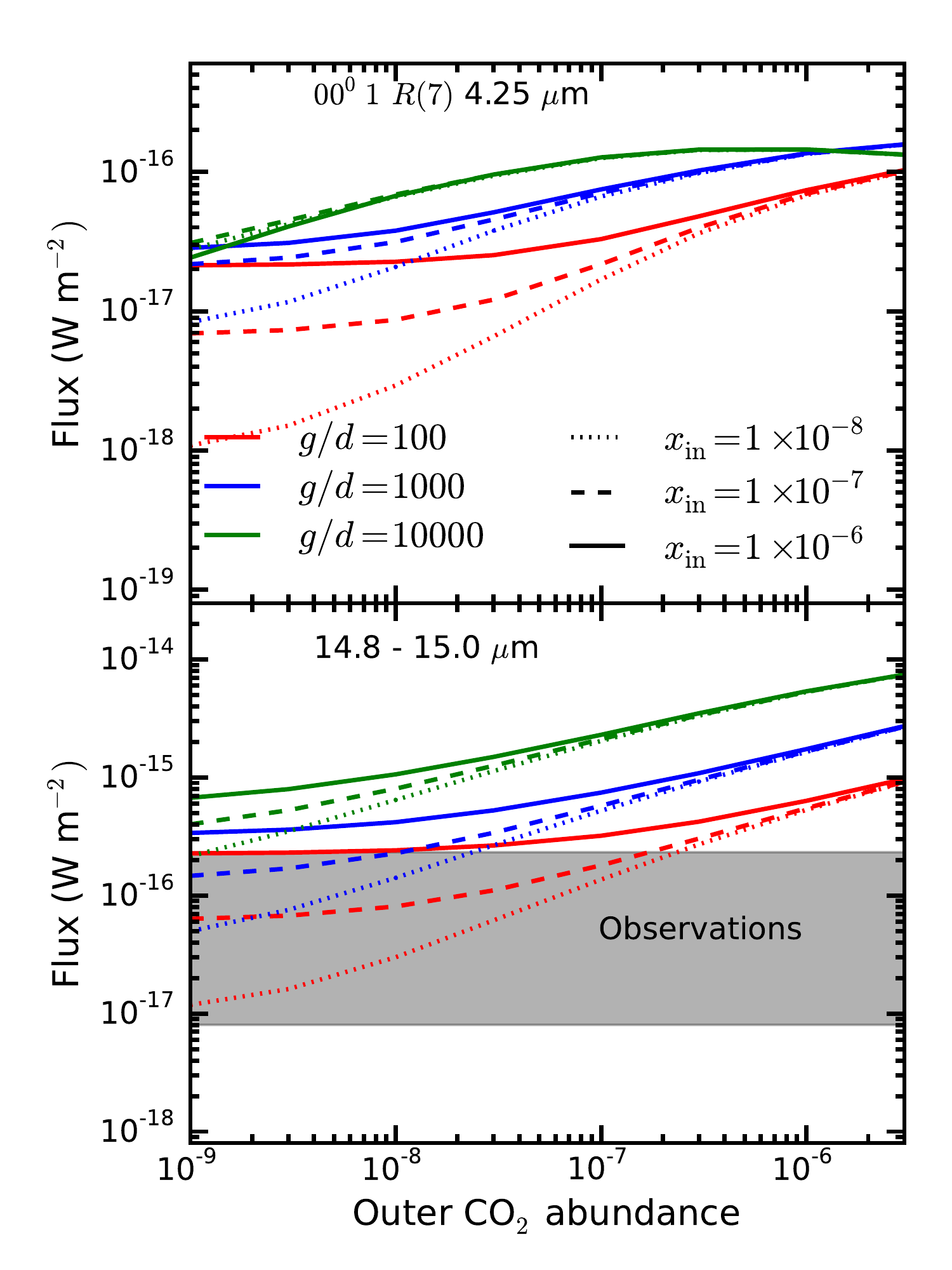}
\caption{\label{fig:Flux_gas} Line fluxes as functions of outer \ce{CO2} abundances for models with constant dust mass ($g/d_{\mathrm{gas}}$) and varying gas/dust ratios. The upper panel shows the flux of the $R$(7) line from the fundamental asymmetric stretch band at 4.3 $\mu$m. The lower panel shows the flux contained in the 15 $\mu$m $Q$-branch feature. The grey region denotes the full range in \ce{CO2} fluxes from the disks that are reported in \cite{Salyk2011}, scaled to the distance of AS 205 (N). The 15 $\mu$m feature contains the flux from multiple $Q-$branches with $\Delta v_2 =1 $. The {\ce{CO2}} flux depends primarily on the outer {\ce{CO2}} abundance and the total g/d ratio and does not strongly depend on the inner {\ce{CO2}} abundances. Only for very low outer {\ce{CO2}} abundances is the effect of the inner abundance on the line fluxes visible. The fluxes for models with $g/d_\mathrm{dust}$ are given in Fig.~\ref{fig:Flux_dust}. }

\end{figure}

Panel $c$ of Fig.~\ref{fig:Gridplotoverview} presents the contribution
function for one of the 15 $\mu$m lines, the $v_2$ $1\rightarrow 0$
$Q(6)$ line. The contribution function shows the relative, azimuthally integrated contribution
to the total integrated line flux. Contours show the areas in which 25\% and 75\% of the
emission is located. Panel $c$ also includes the $\tau = 1$ surface
for the continuum (blue) due to the dust, the $\tau = 1$ surface for
the $v_2$ $1\rightarrow 0$ $Q(6)$ line (red) and surface where the
density is equal to the critical density. The area of the disk
contributing significantly to the emission is large, an annulus from
approximately 0.7 to 30 AU. The dust temperature in the \ce{CO2} emitting region is
between 100 and 500 K and the \ce{CO2} excitation temperature ranges
from 100--300 K (see Fig.~\ref{fig:Overview_app}). The density is lower
 than the critical density at any point in the emitting area.

Panel $d$ of Fig.~\ref{fig:Gridplotoverview} shows the contribution
for the $v_3$ $1\rightarrow0$ $R(7)$ line with the same lines and
contours as panel $c$. The critical density for this line is very
high, $\sim 10^{15}$ cm$^{-3}$. This means that except for the inner 1
AU near the mid-plane, the level population of the $v_3$ level is
dominated by the interaction of the molecule with the surrounding radiation field. The emitting area of
the $v_3$ $1\rightarrow0$ $R(7)$ line is smaller compared to that of
the line at 15 $\mu$m. The emitting area stretches from close the the
sublimation radius up to $\sim 10$ AU. The excitation temperatures for
this line are also higher, ranging from 300--1000 K in the emitting
region (see Fig.~\ref{fig:Overview_app}). 

In Fig.~\ref{fig:Flux_gas} the total flux for the $00^01-00^00\, R(7)$ line
at 4.25 $\mu$m and the 15 $\mu$m feature integrated from 14.8 to 15.0
$\mu$m are presented as functions of $x_\mathrm{out}$, for different
gas-to-dust ratios and different $x_{\mathrm{in}}$. The 15 $\mu$m flux
shows an increase in flux for increasing total {\ce{CO2}} abundance and
gas-to-dust ratio and so does the line flux of the 4.25 $\mu$m line
for most of the parameter space. The total flux never scales linearly
with abundance, due to different opacity
effects. The dust is optically thick at infrared wavelengths up to ~100 AU, so there will always be a reservoir
of gas that will be hidden by the dust. The lines themselves are
strong (have large Einstein $A$ coefficients) and the natural line
width is relatively small (0.2 km s$^{-1}$ FWHM). As a result the
line centers of transitions with low $J$ values quickly become
optically thick. Therefore, if the abundance, and thus the column, in the
upper layers of the disk is high, the line no longer probes the inner
regions. This can be seen in Fig.~\ref{fig:Flux_gas} as the fluxes for
models with different $x_{\mathrm{in}}$ converge with increasing
$x_{\mathrm{out}}$. Convergence happens at lower $x_\mathrm{out}$ for
higher gas-to-dust ratios. The inner region is quickly invisible
through the 4.25 $\mu$m line with increasing gas-to-dust ratios: for a
gas-to-dust ratio of 10000, there is a less than 50\% difference in
fluxes between the models with different inner abundances, even for
the lowest outer abundances. This is not seen so strongly in the 15
$\mu$m feature as it also includes high $J$ lines which are stronger
in the hotter inner regions and are not as optically thick as the low
$J$ lines. There is no significant dependence of the flux on the inner
abundance of {\ce{CO2}} if the outer abundance is $>3\times10^{-7}$ and
the gas to dust ratio is higher than 1000. In these models the 15
$\mu$m feature traces part of the inner 1 AU but only the upper
layers.

Different ways of modelling the gas-to-dust ratio has little effect on
the resulting fluxes. Fig.~\ref{fig:Flux_gas} shows the fluxes for a
constant dust mass and increasing gas mass for increasing the
gas-to-dust ratio, whereas Fig.~\ref{fig:Flux_dust} in Appendix~\ref{app:gd_dust}
shows the fluxes for decreasing dust mass for a constant gas mass.
The differences in fluxes are very small for models with the same gas/dust ratio times \ce{CO2}
abundance, irrespective of the total gas mass: fluxes agree within 10\% for most of the models. This
reflects the fact that the underlying emitting columns of {\ce{CO2}}
are similar above the dust $\tau =1$ surface. Only the temperature of the emitting gas changes: higher temperatures
for gas that is emitting higher up in a high gas mass disk and lower
temperatures for gas that is emitting deeper into the disk in a low
dust mass disk.

The grey band in Fig.~\ref{fig:Flux_gas} and Fig.~\ref{fig:Flux_dust}
shows the range of fluxes observed for protoplanetary disks scaled to
a common distance of 125 pc \citep{Salyk2011}. This figure immediately
shows that low \ce{CO2} abundances, $x_\mathrm{out} < 3\times
10^{-7}$, are needed to be consistent with the observations. Some
disks have lower fluxes than given by the lowest abundance model,
which can be due to other parameters. A more complete comparison
between model and observations is made in Sec.~\ref{ssc:obs_comp}.

In Appendix~\ref{app:LTEvsNLTE} a comparison is made between the
fluxes of models with \ce{CO2} in LTE and models for which the
excitation of \ce{CO2} is calculated from the rate coefficients and
the Einstein $A$ coefficients. The line fluxes differ by a factor of
about three between the models, similar to the differences found by
\cite{Bruderer2015} (their Fig.~6) for the case of HCN.

\subsubsection{The $v_2$ band emission profile}
\label{ssc:15micron}
\begin{figure*}
\centering
\begin{minipage}{0.25\textwidth}
\includegraphics[width=\hsize]{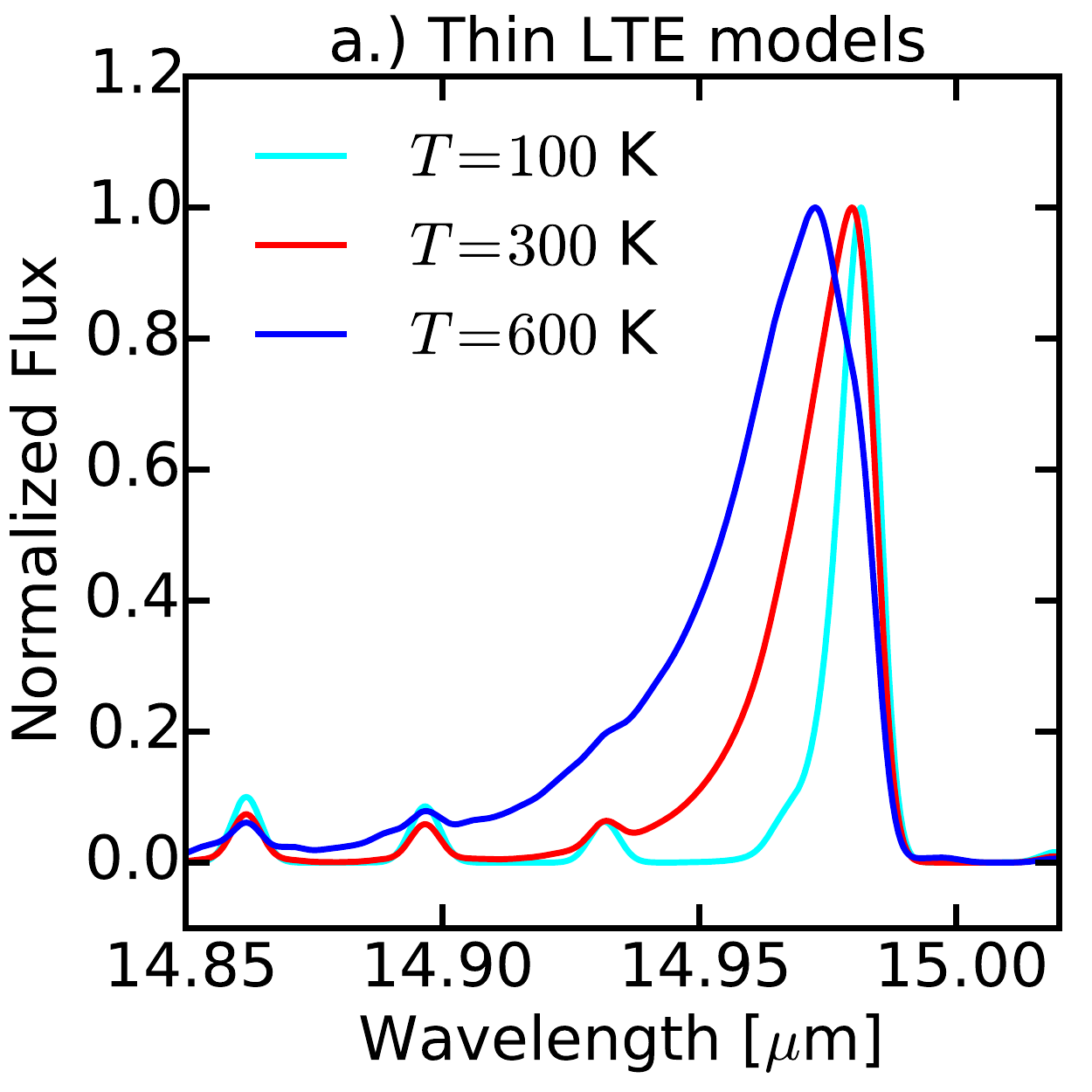}
\end{minipage}%
\begin{minipage}{0.25\textwidth}
\includegraphics[width=\hsize]{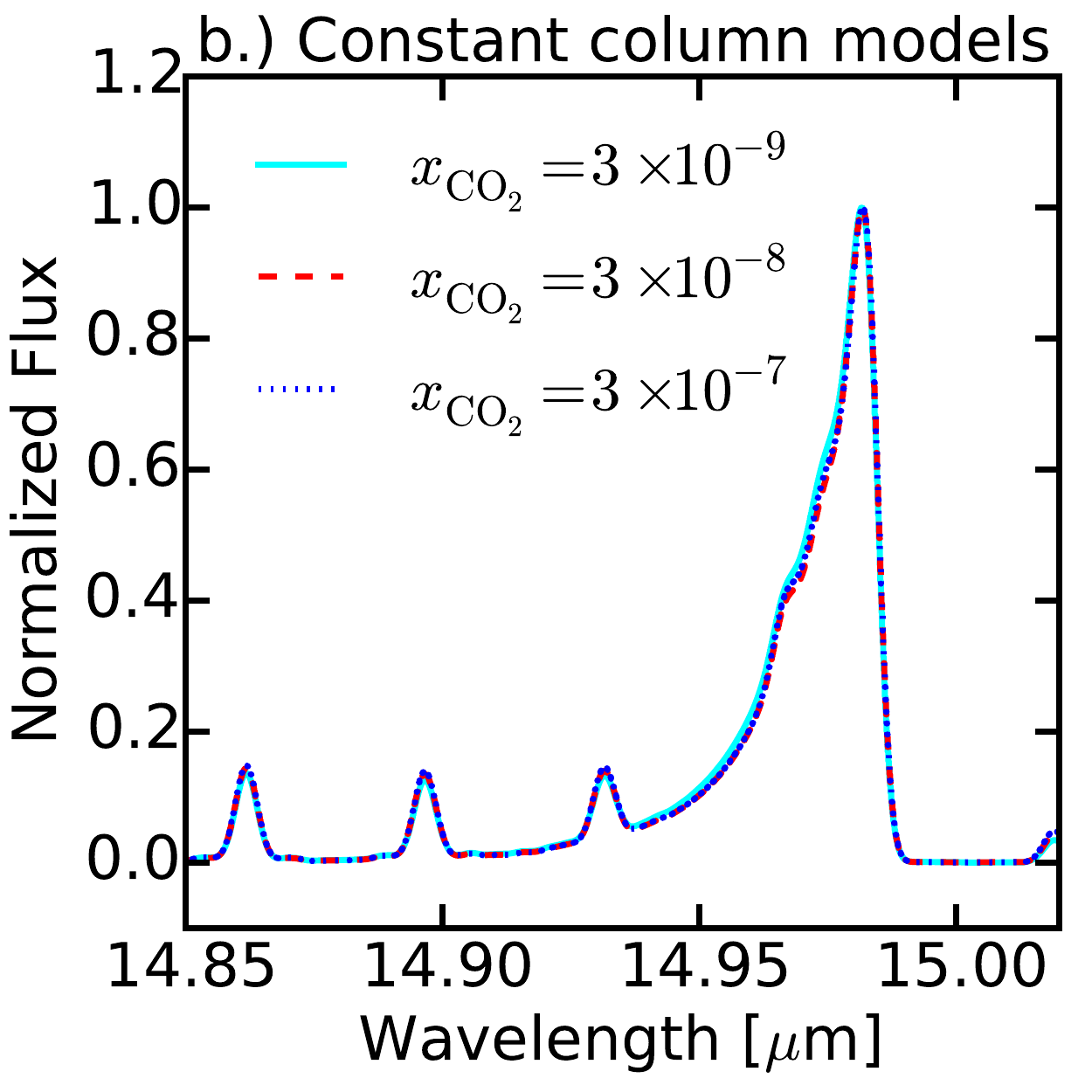}
\end{minipage}%
\begin{minipage}{0.25\textwidth}
\includegraphics[width=\hsize]{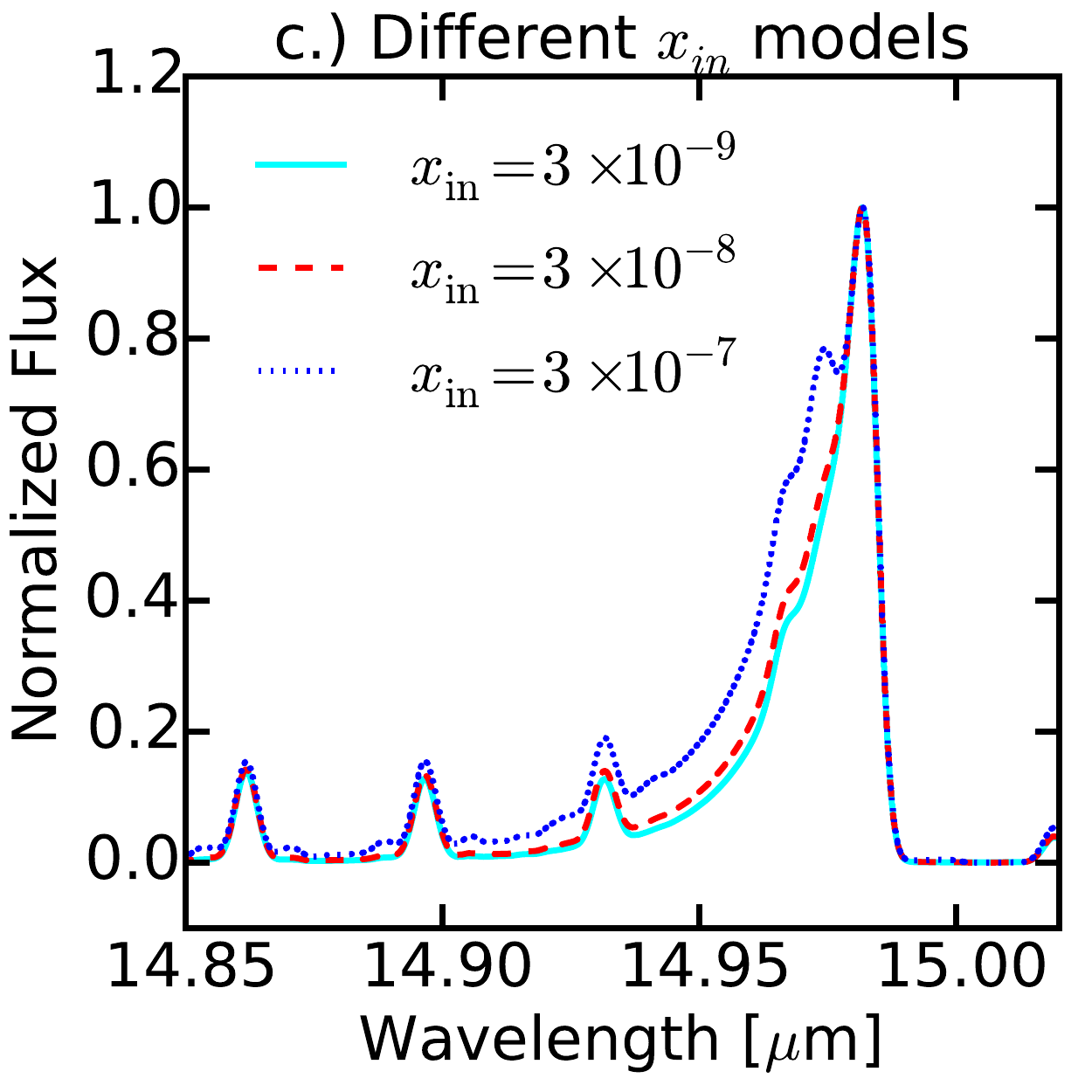}
\end{minipage}%
\begin{minipage}{0.25\textwidth}
\includegraphics[width=\hsize]{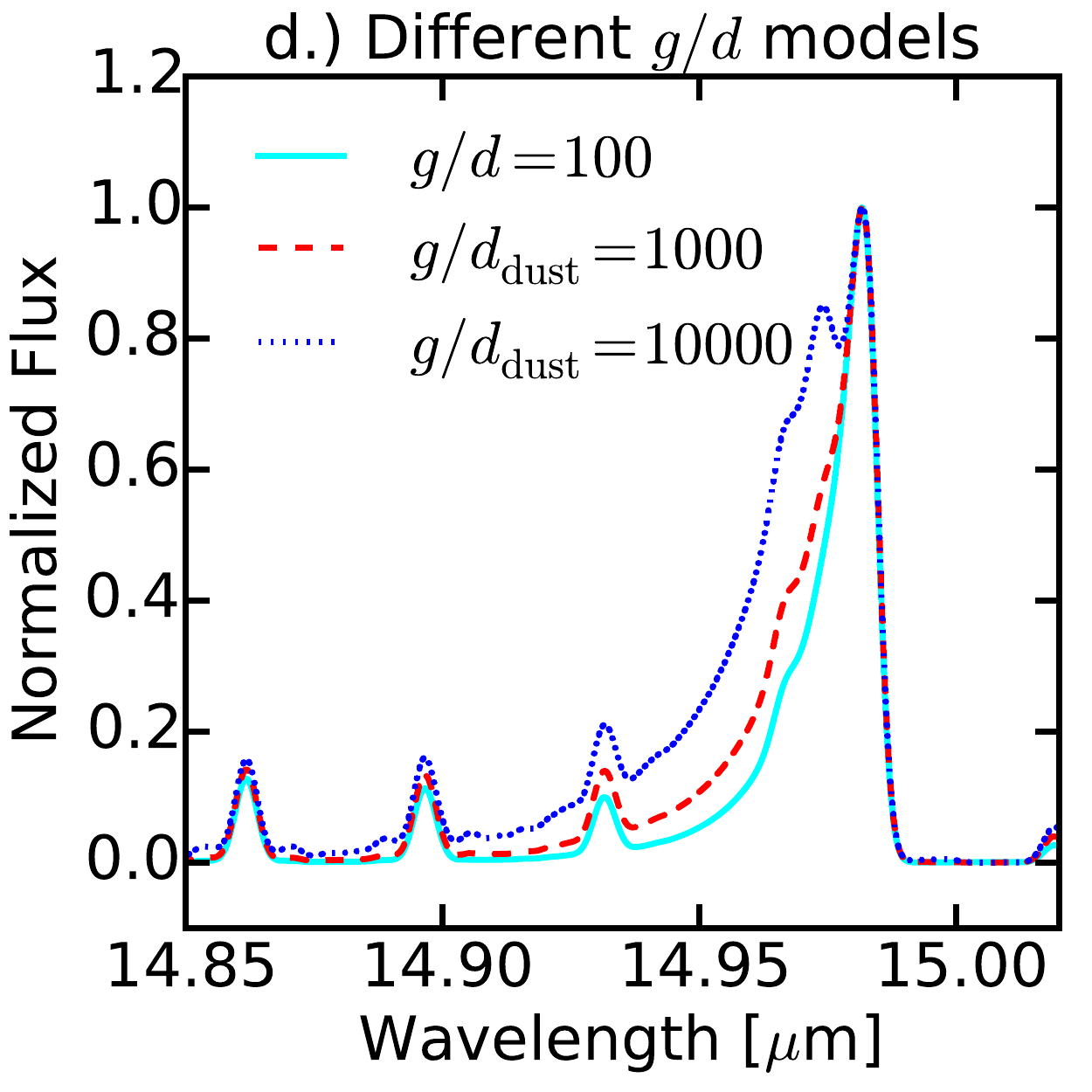}
\end{minipage}
\caption{\label{fig:15microncomparison} $Q$-branch profiles of different models shown at \textit{JWST}-MIRI resolving power. All fluxes are normalized to the maximum of the feature. In the left panel LTE point models with a temperature of 200K (cyan), 400 K (red) and 800 K (blue) are shown. Panel b. shows DALI disk models with a constant abundance profile for which the product of abundance times gas-to-dust ratio is constant.
All these models have very similar total fluxes. The models shown are $g/d_\mathrm{gas} = 100$, $x_{\ce{CO2}} = 3\times10^{-7}$ in red; $g/d_\mathrm{gas}=1000$, $x_{\ce{CO2}} = 3\times 10^{-8}$ in blue and $g/d_\mathrm{gas}=10000$, $x_{\ce{CO2}} = 3\times 10^{-9}$ in cyan. The spectra are virtually indistinguishable.
Panel c. shows DALI disk models with a jump abundance profile, a $g/d_\mathrm{dust}=1000$, an outer \ce{CO2} abundance of $3\times 10^{-8}$ and an inner abundance of $3\times 10^{-7}$ (red), $3\times 10^{-8}$ (blue), $3\times 10^{-9}$ (cyan). The model with the highest inner abundance shows a profile that is slightly broader than those of the other two.
Panel d. shows DALI disk models with the same, constant abundance of $x_{\ce{CO2}} = 3\times 10^{-8}$, but with different $g/d_\mathrm{dust}$ ratios. Removing dust from the upper layers of the disk preferentially boosts the high $J$ lines in the tail of the feature as emission from the dense and hot inner regions of the disk is less occulted by dust.}
\end{figure*}

Fig.~\ref{fig:15microncomparison} shows the $v_2$ $Q$-branch profile
at 15 $\mu$m for a variety of models. All lines have been convolved to
the resolving power of \textit{JWST}-MIRI at that wavelength \citep[$R = 2200$,][]{MIRI2015,Wells2015} with three bins per
resolution element. Panel $a$ shows the results from a simple LTE slab
model at different temperatures whereas panels $b$ and $c$ presents
the same feature from the DALI models. 
Panel $b$ contains models with different gas-to-dust ratios and abundances (assuming $x_{\mathrm{in}} = x_{\mathrm{out}}$) scaled so $g/d \times x_{\ce{CO2}}$ is constant. It shows that gas-to-dust ratio and abundance are degenerate. It is expected that these models show similar spectra, as the total amount of \ce{CO2} above the dust photosphere is equal for all models. The lack of any significant difference shows that collisional excitation of the vibrationally excited state is insignificant compared to radiative pumping.
Panel $c$ of Fig.~\ref{fig:15microncomparison} shows the effect
of different inner abundances on the profile. For the highest inner
abundance shown, $1\times 10^{-6}$, an increase in the shorter
wavelength flux can be seen, but the differences are far smaller than
the differences between the LTE models.
Panel $d$ shows models with similar abundances, but with increasing $g/d_{\textrm{dust}}$. The flux in the 15 $\mu$m feature increases with $g/d_{\textrm{dust}}$ for these models as can be seen in Fig.~\ref{fig:Flux_dust}. This is partly due to the widening of the feature as can be seen in Panel $d$ which is caused by the removal of dust. Due to the lower dust photosphere it is now possible for a larger part of the inner region to contribute to this emission. The inner region is hotter and thus emits more toward high $J$ lines causing the $Q-$branch to widen.

Fitting of LTE models to DALI model spectra  in Fig. 7b-d results in inferred temperatures of 300--600 K. Only the models with a strong tail (blue lines in 7b and 7d) need temperatures of 600 K for a good fit, the other models are well represented with $\sim$ 300 K. For comparison, the actual temperature in the emitting layers is 150--350 K (Fig.~\ref{fig:Overview_app}), illustrating that the optically thin model overestimates the inferred temperatures. The proper inclusion of optical depth effects for the lower-$J$ lines lowers the inferred temperatures. This means that care has to be taken when interpreting a temperature from the \ce{CO2} profile. A wide feature can be due to high optical depths or high rotational temperature of the gas.

\begin{figure*}
\centering
\centering
\begin{minipage}{0.5\textwidth}
\includegraphics[width = \hsize]{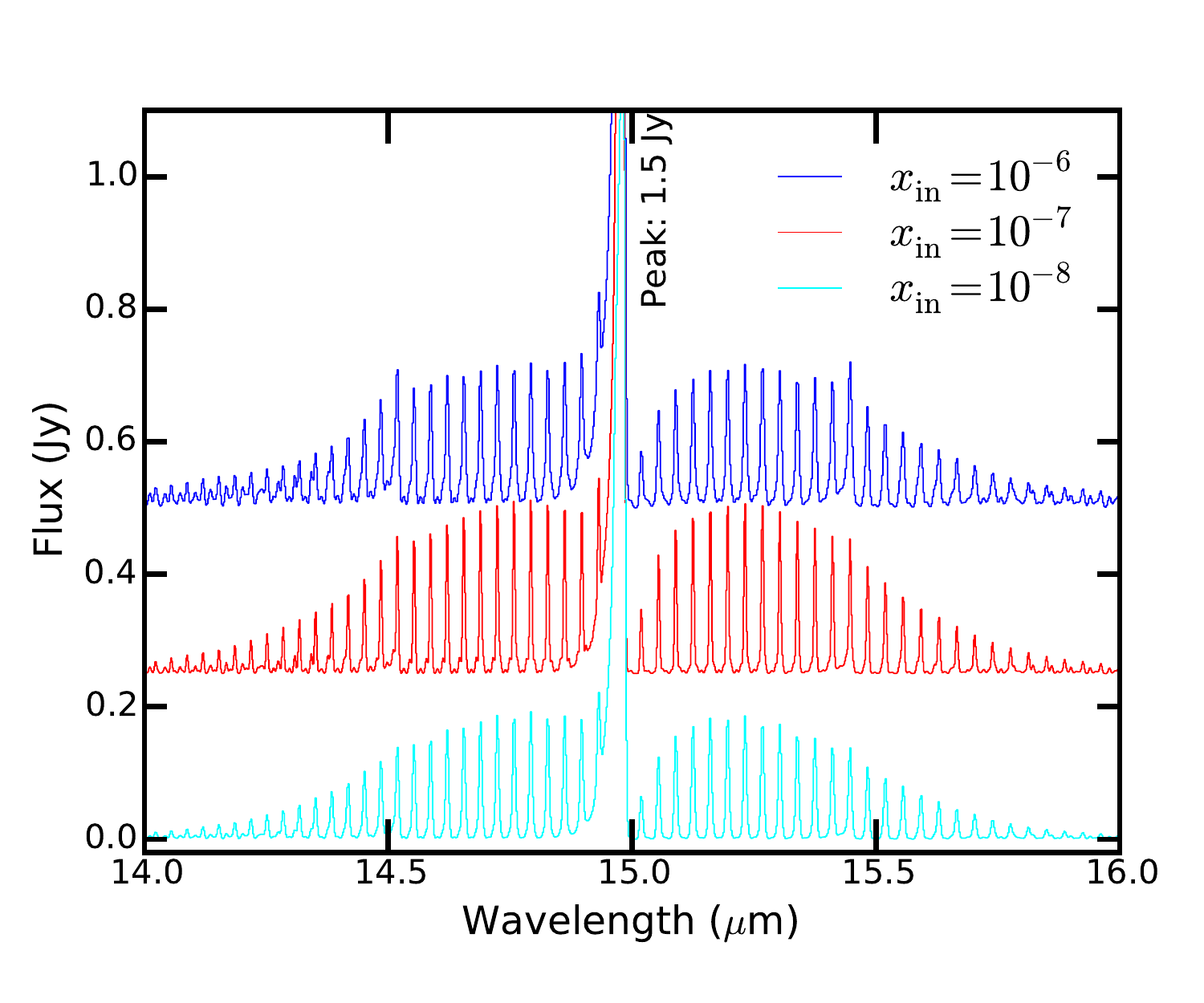}
\end{minipage}%
\begin{minipage}{0.5\textwidth}
\includegraphics[width = \hsize]{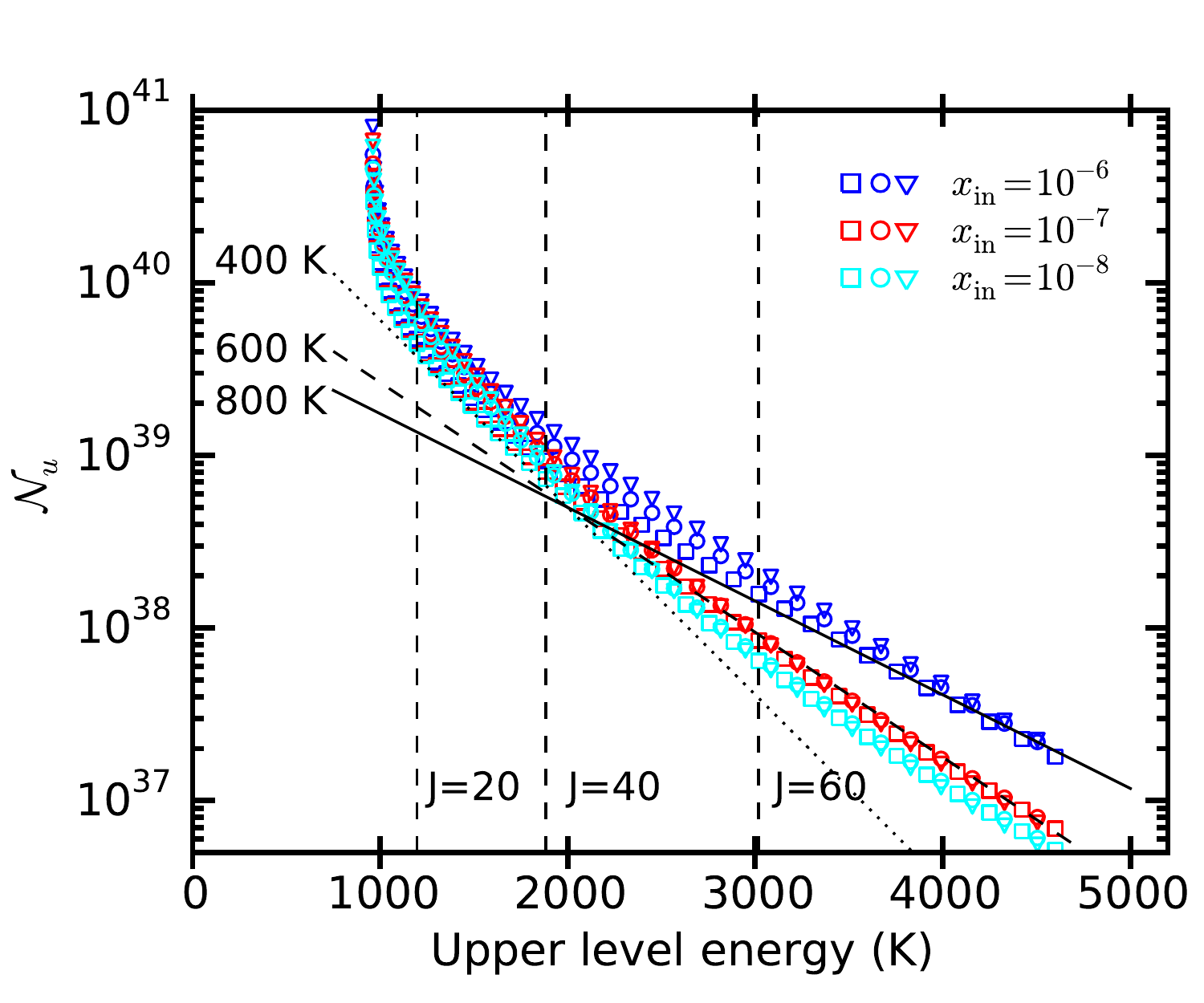}
\end{minipage}
\caption{\label{fig:fulldisk_spectra} Left: Full disk spectra at \textit{JWST}-MIRI resolving power ($R$=2200) for three disk models with different inner {\ce{CO2}} abundances. The outer {\ce{CO2}} abundance is $10^{-7}$ with $g/d_{gas} = 1000$. The models with an inner abundance of $10^{-8}$ and $10^{-7}$ are hard to distinguish, with very similar $P$ and $R-$branch shapes. The spectrum of the model with high inner abundances of $10^{-6}$ are flatter in the region from 14.6 to 14.9 $\mu$m and the wings are also more extended leading to higher high to mid $J$ line ratios. 
Right: Number of molecules in the upper state as function of the upper level energy inferred from the spectra on the left (Boltzmann plot). Inverse triangles denote the number of molecules inferred from $P$-branch lines, squares from $Q$-branch lines and circles from $R$-branch lines. Vertical dashed lines show the upper level energies of the $J = 20,40,60, v2 = 1$ levels. the black dotted, dashed and solid lines show the expected slope for a rotational excitation temperature of 400, 600 and 800 K respectively. The near vertical asymptote near upper level energies of 1000 K (the $v_2 = 1$ rotational ground state energy is due to the regions with large optical depths that dominate the emission from these levels. From around $J = 20$ the curve flattens somewhat and between $J=20$ and $J=40$ the curve is well approximated by the theoretical curve for emission from a 400 K gas. At higher $J$ levels, the model with the highest inner abundance starts to deviate from the other two models as inner and deeper region become more important for the total line emission. Above $J = 60$ the models in with an inner abundance of $10^{-8}$ and $10^{-7}$ are well approximate with a 600 K gas, while the higher inner abundance model is better approximated with a 800 K gas.} 
\end{figure*}

A broader look at the \ce{CO2} spectrum is thus
needed. The left panel of Fig.~\ref{fig:fulldisk_spectra} shows the $P$, $Q$ and
$R-$branches of the vibrational bending mode transition at $R = 2200$,
for models with different inner {\ce{CO2}} abundances and the same
outer abundance of $10^{-7}$. The shape for
the $R-$ and $P-$branches is flatter for
low to mid-$J$ and slightly more extended at high $J$ in the spectrum from the model with an inner
{\ce{CO2}} abundance of $10^{-6}$ than the other
spectra. The peaks at 14.4 $\mu$m and 15.6 $\mu$m are due to the $Q-$branches
from the transitions between $11^10(1) \rightarrow 10^01(1)$ and $11^10(2)
\rightarrow 10^01(2)$ respectively. These are overlapping with lines 
from the bending fundamental P and R branches. For the
constant and low inner {\ce{CO2}} abundances, $10^{-7}$ and $10^{-8}$ respectively $R-$ and $P-$branch shapes are similar,
with models differing only in absolute flux. Decreasing the inner
{\ce{CO2}} abundance from $10^{-8}$ to lower values has no effect of
the line strengths. 

The right panel of Fig.~\ref{fig:fulldisk_spectra} shows Boltzmann plots of the spectra on the left. The number of molecules in the upper state inferred from the flux is given as a function of the upper state energy. The number of molecules in the upper state is given by: $\mathcal{N}_u = 4 \pi d^2F/\left(A_{ul}h\nu_{ul}g_{u}\right)$, with $d$ the distance to the object, F the integrated line flux, $g_u$ the statistical weight of the upper level and $A_{ul}$ and $\nu_{ul}$ the Einstein $A$ coefficient and the frequency of the transition. From slope of $\log(\mathcal{N}_u)$ vs $E_{\textrm{up}}$ a rotational temperature can be determined. The expected slopes for 400, 600 and 800 K are given in the figure. It can be seen that the models do not show strong differences below $J = 20$, where emission is dominated by optically thick lines. Toward higher $J$, the model with $x_{\textrm{in}} = 10^{-6}$ starts to differ more and more from the other two models. The models with $x_{\textrm{in}}=10^{-7}$ and $x_{\textrm{in}}=10^{-8}$ stay within a factor of 2 of each other up to $J = 80$ where the molecule model ends.

Models with similar absolute abundances of
{\ce{CO2}} (constant $g/d \times x_{\ce{CO2}}$) but different
$g/d_{\mathrm{gas}}$ ratios are nearly identical: the width of the $Q$-branch and the
shapes of the $P-$ and $R-$branches are set by the gas temperature
structure. This temperature structure is the same for models with
different $g/d_{\mathrm{gas}}$ ratios as it is set by the dust
structure. The temperature is, however, a function of
$g/d_{\mathrm{dust}}$, but those temperature differences are not large
enough for measurable effects. From this it also follows that the exact collisional rate coefficients are not important:
The density is low enough that the radiation field can set the excitation of the vibrational levels. At the same time the density is still high enough to be higher than the critical density for the rotational transitions, setting the rotational excitation temperature equal to the gas kinetic temperature.   

The branch shapes are a function of $g/d_{\mathrm{dust}}$ at constant
\textit{absolute} abundance. Apart from the total flux which is slightly higher at higher
$g/d_{\mathrm{dust}}$  (Fig.~\ref{fig:Flux_dust}), the spectra are also broader (Panel d. Fig.~\ref{fig:15microncomparison}). This
is because the hotter inner regions are less occulted by dust for higher $g/d_{\mathrm{dust}}$ ratios. This hotter gas has more emission coming from high $J$ lines, boosting the tail of the $Q-$branch.

To quantify the effects of different abundance profiles, line ratios
can also be informative. The lines are chosen so they are free from
water emission (see Appendix~\ref{app:water}). The top two panels of
Fig.~\ref{fig:Lineratios} shows the line ratios for lines in the
$01^10(1) \rightarrow 00^00(1)$ 15 $\mu$m band: $R(37):R(7)$ and
$P(15):P(51)$. The $R(7)$ and $P(15)$ lines come from levels with
energies close to the lowest energy level in the vibrational state
(energy difference is less than 140 K). These levels are thus easily
populated and the lines coming from these levels are quickly optically
thick. The $R(37)$ and $P(51)$ lines come from levels with rotational
energies at least 750 K above the ground vibrational energy. These
lines need high kinetic/rotational temperatures to show up strongly
and need higher columns of {\ce{CO2}} at prevailing temperatures to
become optically thick. From Fig.~\ref{fig:Lineratios} a few things
become clear. First, for very high outer abundances, it
is very difficult to distinguish between different inner abundances
based on the presented line ratio. Second, models with high outer abundances
are nearly degenerate with models that have a low outer abundance and
a high inner abundance. A measure of the optical depth will solve
this. In the more intermediate regimes the line ratios presented here
 or a Boltzmann plot will supplement the information needed to distinguish
 between a cold, optically thick \ce{CO2} reservoir and a hot, more optically 
 thin \ce{CO2} reservoir that would be degenerate in just $Q-$branch fitting.

\begin{figure}
\centering
\includegraphics[width = \hsize]{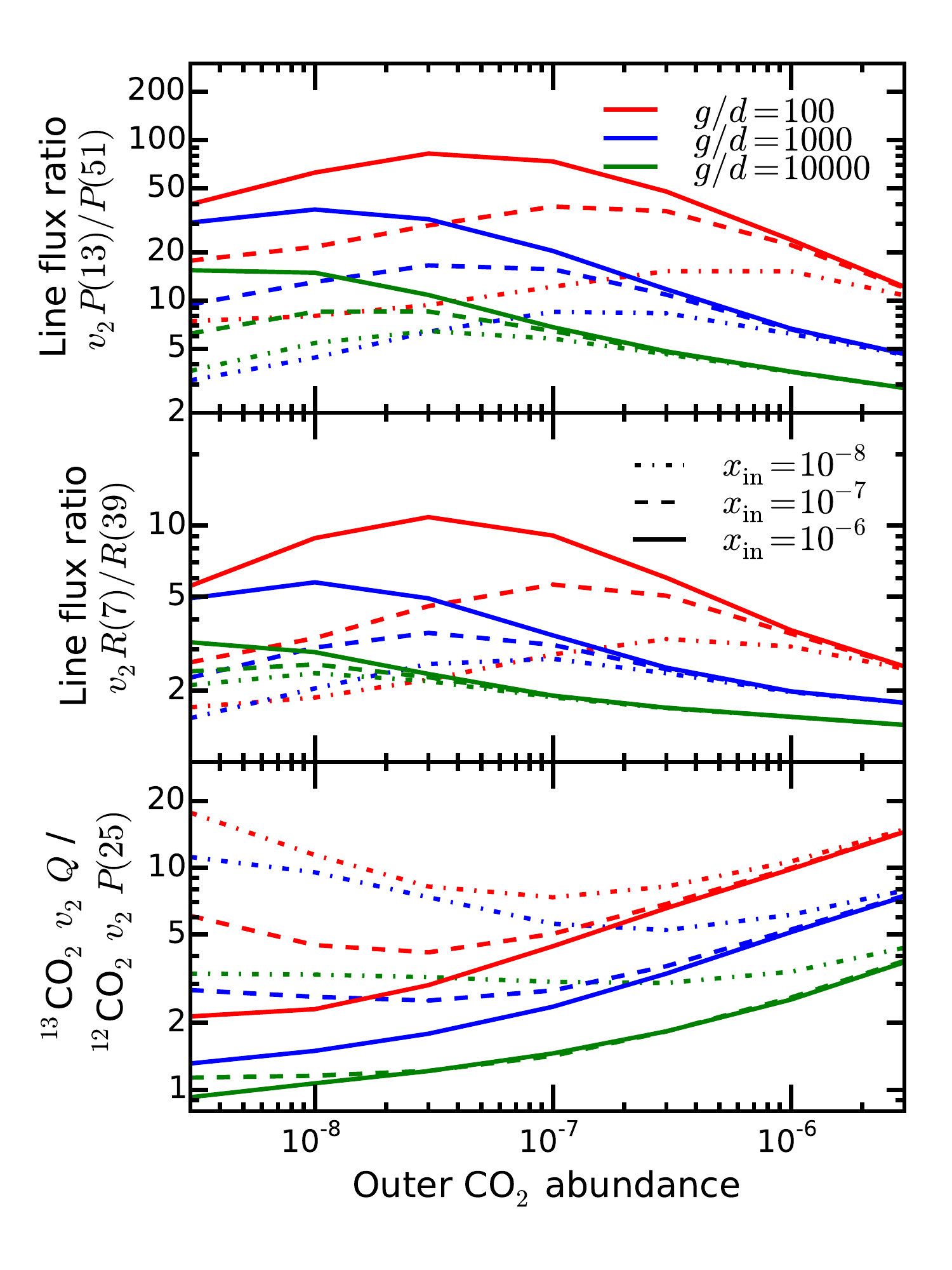}
\caption{\label{fig:Lineratios} Line ratios as functions of outer abundance, inner abundance and gas to dust ratio ($g/d_{gas}$). One line ratio in the P branch ($P(15)$:$P(51)$) (top panel), one line ratio in the R branch ($R(7)$:$R(39))$ (middle panel) of the $01^10 \rightarrow 00^00$ 15 $\mu$m transition are shown together with the line ratio between the \ce{^{13}CO2} $Q-$branch and the neighbouring \ce{^{12}CO2} $P(25)$ line. See the main text for more details.
}
\end{figure}

\subsubsection{\ce{^{13}CO2} $v_2$ band}
\label{ssc:13CO2}
An easier method to break these degeneracies is to use the \ce{^{13}CO2}
isotopologue. \ce{^{13}CO2} is approximately 68 times less abundant
compared to \ce{^{12}CO2}, using a standard local interstellar medium
value \citep{Wilson1994,Milam2005}. This means that the isotopologue is much less likely
to be optically thick and thus \ce{^{13}CO2}:\ce{^{12}CO2} line ratios
can be used as a measure of the optical depth, adding the needed
information to lift the degeneracies. The bottom panel of
Fig.~\ref{fig:Lineratios} shows the ratio between the flux in the
\ce{^{13}CO2} $v_2$ $Q$-branch and the \ce{^{12}CO2} $v_2$ $P(25)$
line.

\begin{figure}
\centering
\includegraphics[width = \hsize]{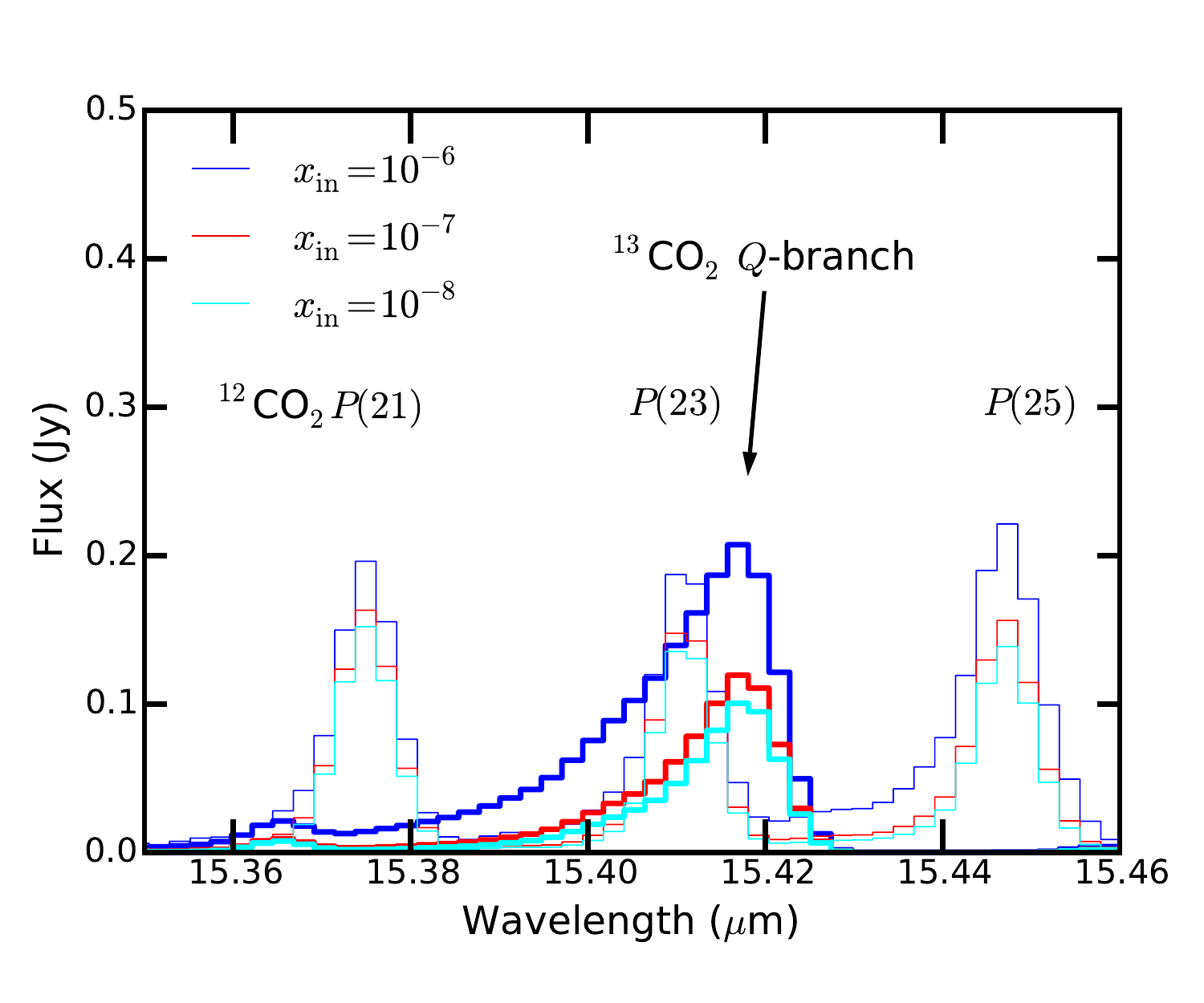}
\caption{\label{fig:13CO2_Q} Zoom in of Fig.~\ref{fig:fulldisk_spectra} with \ce{^{13}CO2} emission added to the spectra (thick lines). The \ce{^{13}CO2} $Q-$branch is more sensitive to higher inner abundances.  } 
\end{figure}

As the $Q-$branch for \ce{^{13}CO2} is less optically thick, it is
also more sensitive to the abundance structure. The $Q-$branch,
situated at 15.42 $\mu$m, partially overlaps with the $P(23)$ line of the
more abundant isotopologue so both isotopologues need to be modelled
to properly account the the contribution of these
lines. Fig.~\ref{fig:13CO2_Q} shows the same models as in
Fig.~\ref{fig:fulldisk_spectra} but now with the \ce{^{13}CO2}
emission in thick lines. The \ce{^{13}CO2} Q-branch is predicted to be
approximately as strong as the nearby \ce{^{12}CO2} lines for the
highest inner abundances. The total flux in the \ce{^{13}CO2}
$Q$-branch shows a stronger dependence on the inner {\ce{CO2}}
abundance than the \ce{^{12}CO2} $Q$-branch. A hot reservoir of
{\ce{CO2}} strongly shows up as an extended tail of the \ce{^{13}CO2}
$Q$-branch between 15.38 and 15.40 $\mu$m.

\subsubsection{Emission from the $v_3$ band}
\begin{figure}
\centering
\includegraphics[width = \hsize]{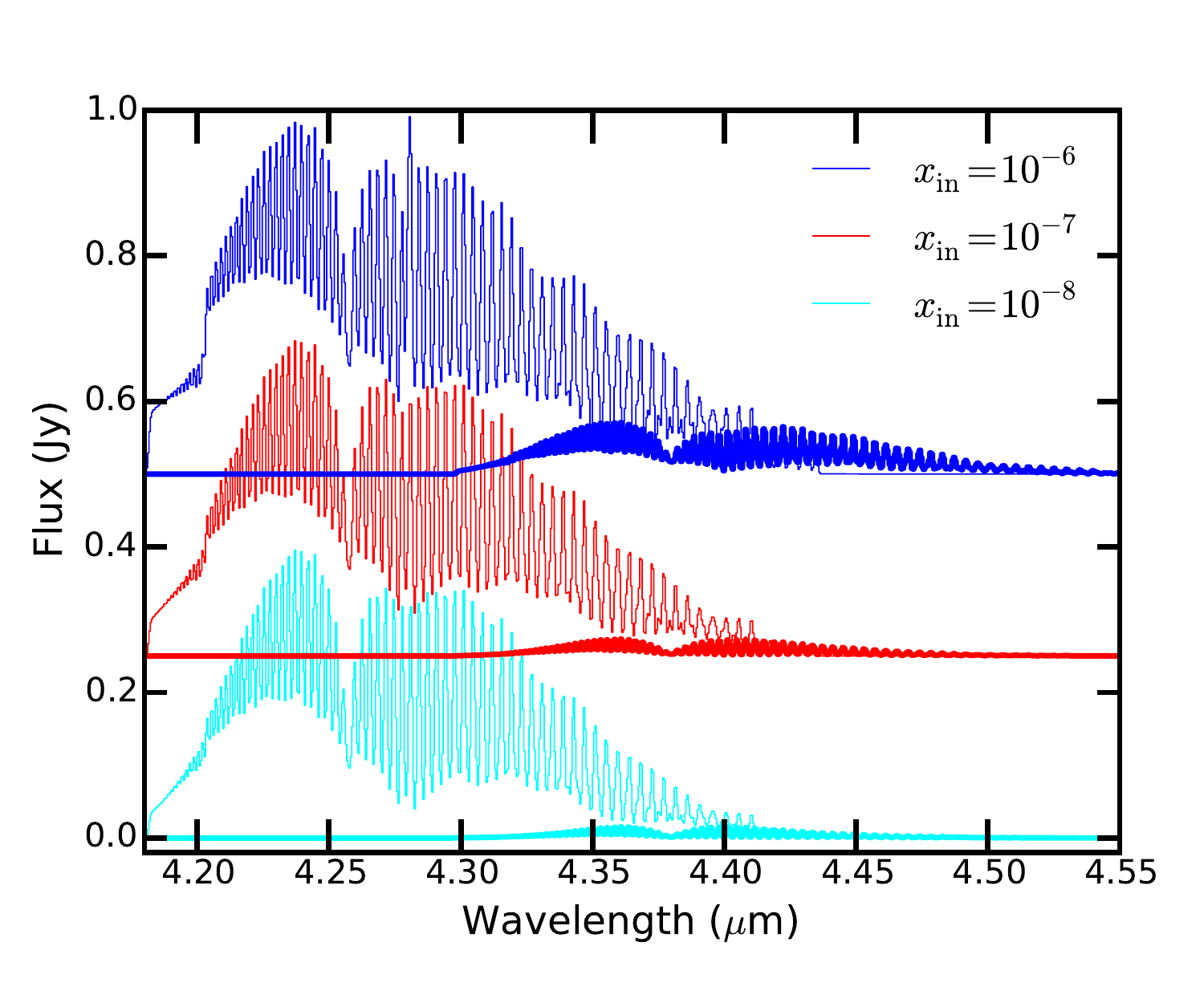}
\caption{\label{fig:4.3micronspectrum} Full disk 4.3 $\mu$m spectra at
  $R$=3000 for three disk models with different inner {\ce{CO2}}
  abundances. The outer \ce{^{12}CO2} abundance is $10^{-7}$ with
  $g/d_{gas} = 1000$. Thin lines show the emission from \ce{^{12}CO2}, thick
  lines the emission from \ce{^{13}CO2}. For both isotopes individual line peaks can be seen but the lines
  blend together in the wings forming a single band. Spectra are shifted vertically for display purposes.}
\end{figure}

The $v_3$ band around 4.25 $\mu$m is a strong emission band in the
disk models, containing a larger total flux than the $v_2$ band. Even
so, the 4.3 $\mu$m band of gaseous \ce{CO2} has not been seen in
observations of \textit{ISO} with the Short Wave Spectrometer (SWS) toward high mass protostars in contrast with
15 $\mu$m band that has been seen towards these sources in absorption
\citep{Dishoeck1996,Boonman2003CO2}. This may be largely due to the
strong solid \ce{CO2} 4.2 $\mu$m ice feature obscuring the gas-phase
lines for the case of protostars, but for disks this should not be a
limitation. Fig.~\ref{fig:4.3micronspectrum} shows the spectrum of
gaseous {\ce{CO2}} in the $v_3$ band around 4.3 $\mu$m at
\textit{JWST}-NIRSpec resolving power. The resolving power of NIRSpec
is taken to be $R=3000$, which is not enough to fully separate the
lines from each other. The {\ce{CO2}} emission thus shows up as an
extended band.

The band shapes in Fig.~\ref{fig:4.3micronspectrum} are very
similar. The largest difference is the strength of the 4.2 $\mu$m
discontinuity, which is probably an artefact of the model as only a finite
number of $J$ levels are taken into account. The total flux over the
whole feature does depend on the inner abundance, but the difference
is of the order of $\sim10\%$ for 2 orders of magnitude change of the
inner abundance.

Fig.~\ref{fig:4.3micronspectrum} also shows the \ce{^{13}CO2}
spectrum. The lines from \ce{^{13}CO2} are mostly blended with much
stronger lines from \ce{^{12}CO2}. At the longer wavelength limit,
\ce{^{13}CO2} lines are stronger than those of \ce{^{12}CO2} but there
the 6 $\mu$m water band and 4.7 $\mu$m \ce{CO} band start to complicate
the detection of \ce{^{13}CO2} in the 4--5 $\mu$m region.

Average abundances of \ce{CO2} can be derived from observations of the 4.3 $\mu$m band. While inferring the abundance structure will be easier from the 15 $\mu$m band there are some observational advantages of using the 4.3 $\mu$m band. NIRSpec has multi-object capabilities and will thus be able to get large samples of disks in a single exposure, especially for more distant clusters where there are many sources in a single FOV. NIRSpec has the additional advantages that it does not suffer from detector fringing and that it is more sensitive \citep[NIRSpec pocket guide\footnote{\url{https://jwst.stsci.edu/}},][]{MIRI2015,Wells2015}.  As both the 4.3 $\mu$m and 15 $\mu$m bands are pumped by infrared radiation, the flux ratios between these two will mostly contain information about the ratio of the continuum radiation field between the wavelengths of these bands.

\subsection{Line-to-continuum ratio}
\label{ssc:line-cont}
\begin{figure}
\centering
\includegraphics[width = \hsize]{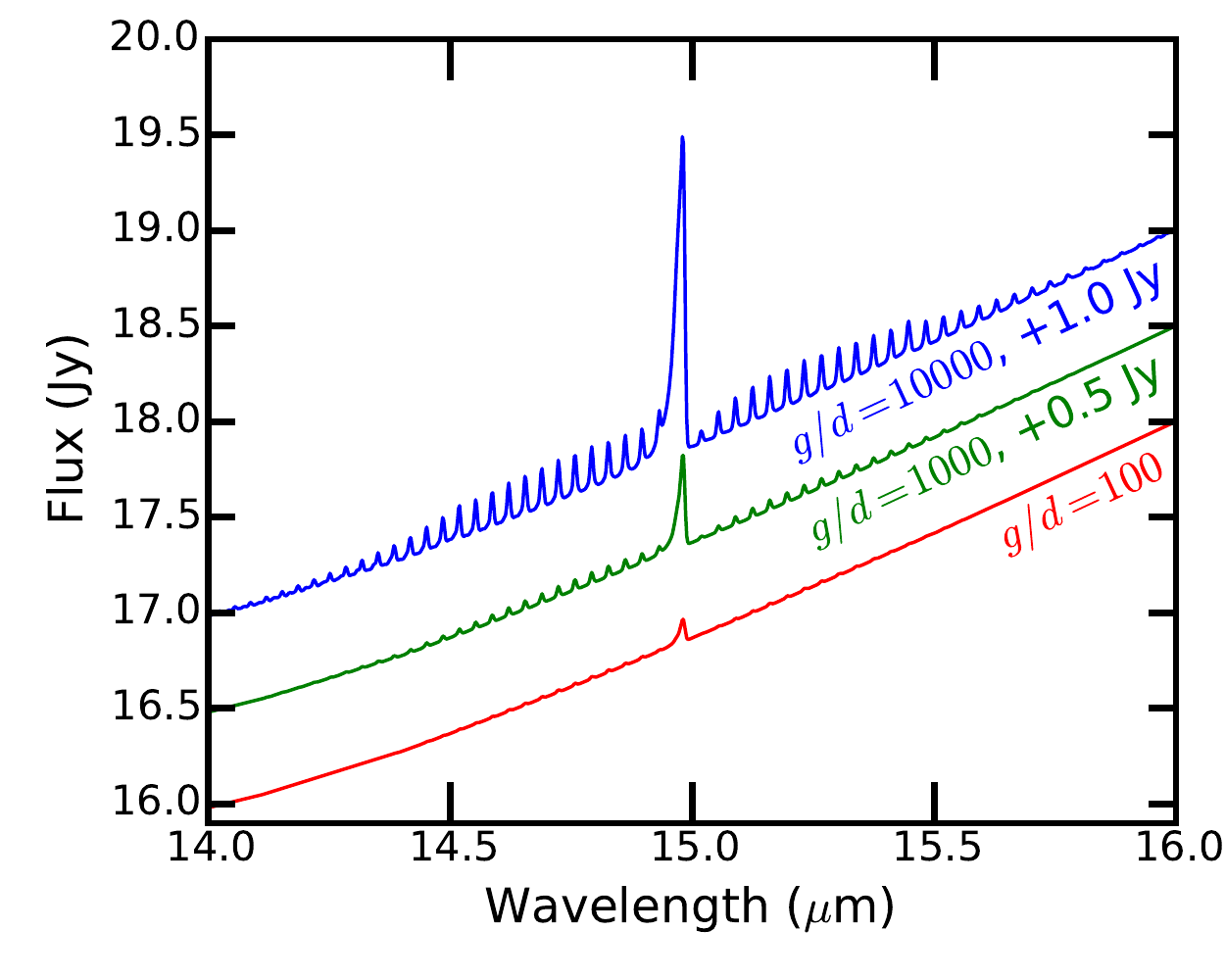}
\caption{\label{fig:continuum_spectra} Full disk spectra with added continuum for models with different gas-to-dust ratios. All spectra have been convolved to a spectral resolving power of $R = 2200$. All models have the same \ce{CO2} abundance of $10^{-8}$. The spectra have been shifted by the amount shown.}
\end{figure}

The line-to-continuum ratio is potentially an even better diagnostic
of the gas/dust ratio than line ratios \citep{Meijerink2009}.
Fig.~\ref{fig:continuum_spectra} presents spectra with the continuum
added to it. The spectra have been shifted, as the continua for these models overlap. The
models for which the spectra are derived all have a \ce{CO2} abundance
of $10^{-8}$ but differ in the gas-to-dust ratio. The
gas-to-dust ratio, determines the column of \ce{CO2}
that can contribute to the line. A large part of the \ce{CO2}
reservoir near the mid-plane cannot contribute due to the large
continuum optical depth of the dust. It is thus not surprising that
the line-to-continuum ratio is strongly dependant on the gas-to-dust
ratio. The precise way of setting the gas-to-dust ratio (by increasing
the amount of gas, or decreasing the amount of dust) does not really
matter for the line-to-continuum ratio. It does matter for the
absolute scaling of the continuum, which decreases if the amount of
dust is decreased. 

\begin{figure}
\centering
\includegraphics[width = \hsize]{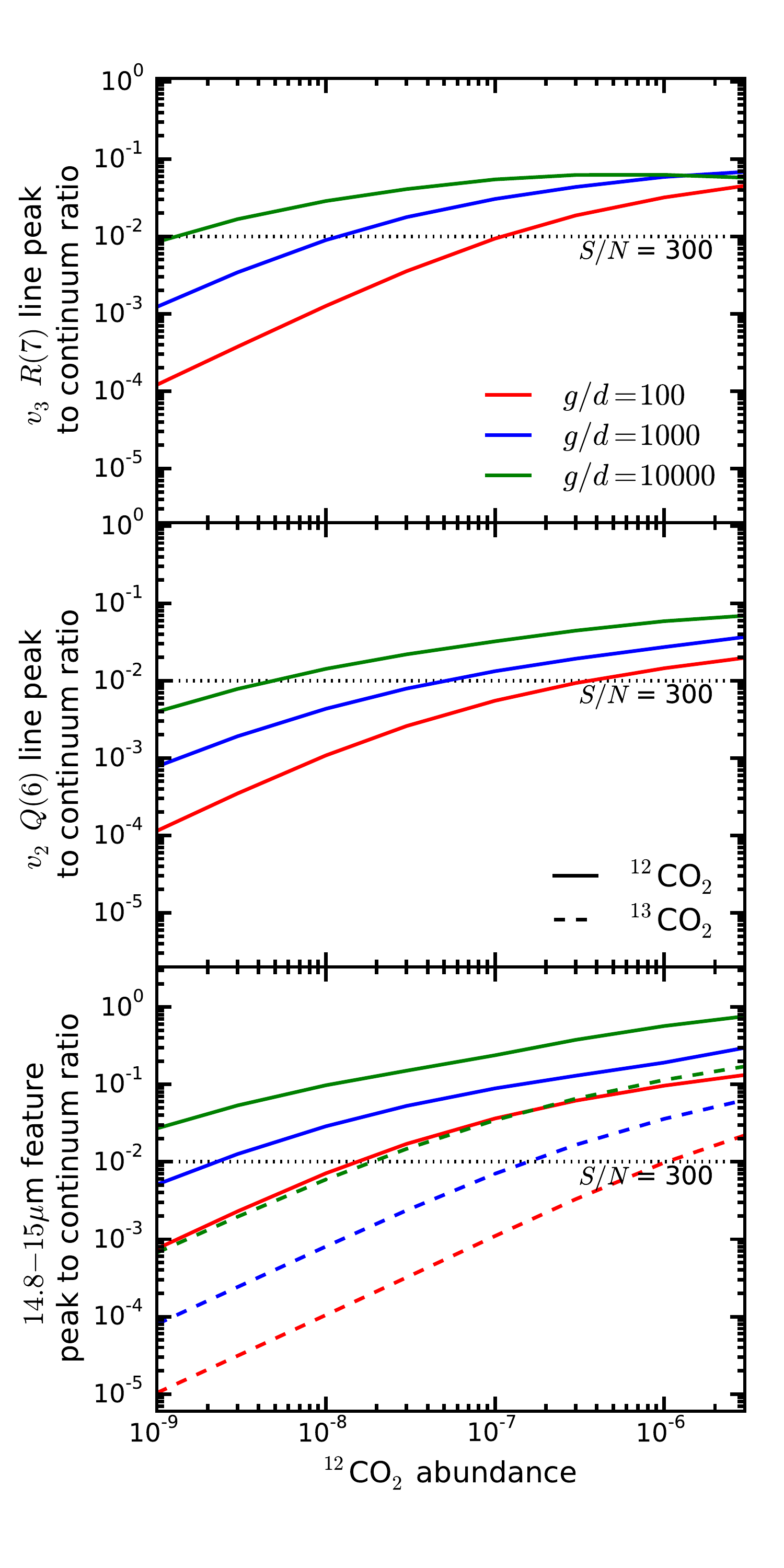}
\caption{\label{fig:line-to-continuumplot_iso} Line-to-continuum for \ce{^{12}CO2} as function of abundance for different gas-to-dust ratios in solid lines. In the bottom plot the line-to-continuum for \ce{^{13}CO2} $Q-$branch is shown in dashed lines. A dotted black line shows a line-to-continuum of 0.01, lines with this line-to-continuum ratio can be observed if the signal-to-noise (S/N) on the continuum is more than 300. With a S/N on the continuum of 300, observations of the $Q-$branch should be able to probe down to $10^{-9}$ in abundance for a gas-to-dust ratio of 1000. With similar gas-to-dust ratio and S/N, individual lines the 15 $\mu$m band will only be observable in disk with \ce{CO2} abundances higher than $3\times 10^{-8}$.}
\end{figure}

\cite{Meijerink2009} could constrain the gas-to-dust ratio from the data since there is an upper limit to the \ce{H2O} abundance from the atomic \ce{O} abundance. Here it is not possible to make a similar statement as \ce{CO2} is not expected to be a major reservoir of either the oxygen or the carbon in the disk. On the contrary, from Fig.~\ref{fig:Flux_gas} it can be seen that with a gas-to-dust ratio of 100 an abundance of $10^{-7}$ is high enough to explain the brightest of the observed line fluxes. The line-to-continuum ratios from the \textit{Spitzer}-IRS spectra of 5--10\% are also matched by the same models (see Fig.~\ref{fig:line-to-continuumplot_iso}). External information such as can be obtained from \ce{H2O} is needed to lift the degeneracy between high gas-to-dust ratio and high abundance: if one of the two is fixed, the other can be determined from the flux or line-to-continuum ratio. 

The line-to-continuum ratio is very important for planning observations, however, as it sets the limit on how precisely the continuum needs to be measured to be able to make a robust line detection. Fig.~\ref{fig:line-to-continuumplot_iso} shows the line-to-continuum ratios for models with a constant abundance. These figures show that high signal-to-noise ($S/N$) on the continuum is needed to be able to get robust line detections. The \ce{^{12}CO2} $Q-$branch should be easily accessible for most protoplanetary disks. To be able to probe individual $P-$ and $R-$branch lines of the \ce{^{12}CO2} 15 $\mu$m feature as well as the \ce{^{13}CO2} $Q-$branch, deeper observations (reaching $S/N$ of at least 300, up to 1000) will be needed to probe down to disks with \ce{CO2} abundances of $10^{-8}$ and gas-to-dust ratios of 1000.

\subsection{{\ce{CO2}} from the ground}
\label{ssc:CO2ground}

As noted earlier, there is a large part of the {\ce{CO2}} spectrum
that cannot be seen from the ground because of atmospheric
\ce{CO2}. There are a few lines, however, that could be targeted from
the ground using high spectral resolution. The high $J$ lines
($J > 70$) of the $v_1=1-0$ transition in the $R$ branch around 4.18
$\mu$m are visible with a resolving power of $R = 30000$ or higher. At
this resolution the {\ce{CO2}} atmospheric lines are resolved and at
$J > 70$ they are narrow enough to leave $20-50\%$ transmission windows
between them (ESO skycalc\footnote{\url{http://www.eso.org/observing/etc/bin/gen/form?INS.MODE=swspectr+INS.NAME=SKYCALC}}). The lines are expected to have a peak line-to-continuum ratio of 1:100. So a $S/N$ of 10 on the line peak translates to a $S/N$ of 1000 on the continuum. The FWHM of the atmospheric lines is about 30 km s$^{-1}$. So half of the emission line profile should be observable when the relative velocity shift between observer and source is more than 15 km s$^{-1}$. Since the Earth's orbit allows for velocity shifts up to 30 km s$^{-1}$ in both directions, observing the full line profile is possible in two observations at different times of year for sources close to the orbital plane of the earth. 
The exposure time needed to get a $S/N$ of 1000 on the continuum at 
4.18 $\mu$m, which is 6.7 Jy in our AS 205 (N) model, with a 40\% sky transmission on the lines, is about 10 hours for VLT-CRIRES \footnote{Exposure times have been calculated with the ESO exposure time calculator \url{https://www.eso.org/observing/etc/} for CRIRES (version 5.0.1)}. The
high $J$ $P$-branch lines of {\ce{CO2}} are close to atmospheric
lines from \ce{N2O}, \ce{O3} and \ce{H2O} resulting in a very opaque atmosphere at these wavelengths \citep{Noll2012,Jones2013}.

The other lines that can be seen from the ground are between 9 and 12
$\mu$m (N-band). These originate from the $01^11$ and $00^01$
levels. The line to continuum ratios vary from 1:40 to 1:3000 for the
brighter lines in the $00^01 \rightarrow 10^00(1)$ band with the most
likely models having line to continuum ratios between 1:200 and
1:2000. For a continuum of $\sim$ 11 Jy a $S/N$ of 2000 for $R=10^5$
could potentially be achieved in about three to ten minutes of integration with the European
Extremely Large Telescope (E-ELT). At the location of the \ce{CO2} atmospheric absorption lines in this part of the spectrum the sky transmittance is $\sim$50\%  and the atmospheric lines have a FWHM of $\sim$ 50 km s$^{-1}$

\subsection{\ce{CO2} model uncertainties}

The fluxes derived from the DALI models depend on the details of the
{\ce{CO2}} excitation processes included in the model. The collisional
rate coefficients are particularly uncertain, since the measured set
is incomplete. There are multiple ways to extrapolate what is measured
to what is needed to complete the model. Modelling slabs of \ce{CO2} 
using different extrapolations such as: absolute scaling of the rotational 
collision rate coefficients, including temperature dependence of the 
vibrational collisional rate coefficients and different implementations 
of the collision rate coefficients between the vibrational levels with $2v_1 + v_2 = \textrm{constant}$, show that
fluxes can change by up to 50\% for specific combinations of radiation
fields and densities. The highest differences are seen in the 4--5
$\mu$m band, usually at low densities. The flux in the 15 $\mu$m band
usually stays within 10\% of the flux of the model used here. These
uncertainties are small compared to other uncertainties in disk
modelling such as the chemistry or parameters of the disk hosting
protostar. The main reason that the fluxes are relatively insensitive
to the details of the collisional rate coefficients is due to the
importance of radiative pumping in parallel with collisions. 

The assumption $T_\textrm{dust} = T_\textrm{gas}$ is not entirely correct, since the gas temperature can be up to 5\% higher than the dust temperature in the \ce{CO2} emitting regions. This affects our line fluxes. For the 4.3 $\mu$m fluxes the induced difference in flux is always smaller than 10\%. For the 15 $\mu$m fluxes difference are generally smaller than 10\%, whereas some of the higher J lines are up to 25\% brighter.

A very simplified abundance structure was taken. It is likely that
protoplanetary disks will not have the abundance structure adopted
here. Full chemical models indeed show much more complex chemical
structures \citep[see e.g.][]{Walsh2015}. The analysis done here
should still hold for more complex abundance structures, and future
work will couple such chemistry models directly with the excitation
and radiative transfer.

The stellar parameters for the central star and the exact parameters
of the protoplanetary disk also influence the resulting \ce{CO2}
spectrum. The central star influences the line emission through its UV
radiation that can both dissociate molecules and heat the gas. Since
for our models no chemistry is included, only the heating of the
dust by stellar radiation is important for our models. The \ce{CO2}
flux in the emission band around 15 $\mu$m scales almost linearly with
the bolometric luminosity of the central object (see
Fig.~\ref{fig:lum_flux}).

\section{Discussion}

\subsection{Observed 15 $\mu$m profiles and inferred abundances}
\label{ssc:obs_comp}

The $v_2$ 15 $\mu$m feature of {\ce{CO2}} has been observed in many
sources with \textit{Spitzer-IRS}
\citep{Pontoppidan2010,Salyk2011}. The SH (Short-High) mode barely
resolves the 15 $\mu$m $Q$-branch, but that is enough to compare with the
models. We used the spectra that have been reduced with the Caltech
High-res IRS pipeline (CHIP) \citep{Pontoppidan2010,CHIP2016}. The
sources selected out of the repository have a strong emission feature
of {\ce{CO2}} but no distinguishable \ce{H2O} emission in the 10--20
$\mu$m range. The sources and some stellar parameters are listed in
Table~\ref{tab:stellarparam}. The observed spectra are continuum
subtracted (Appendix~\ref{app:spi_spec}) and the observed profiles are
compared with model profiles by eye (Fig.~\ref{fig:cont_sub}). 

\begin{figure}
\centering
\includegraphics[width=\hsize]{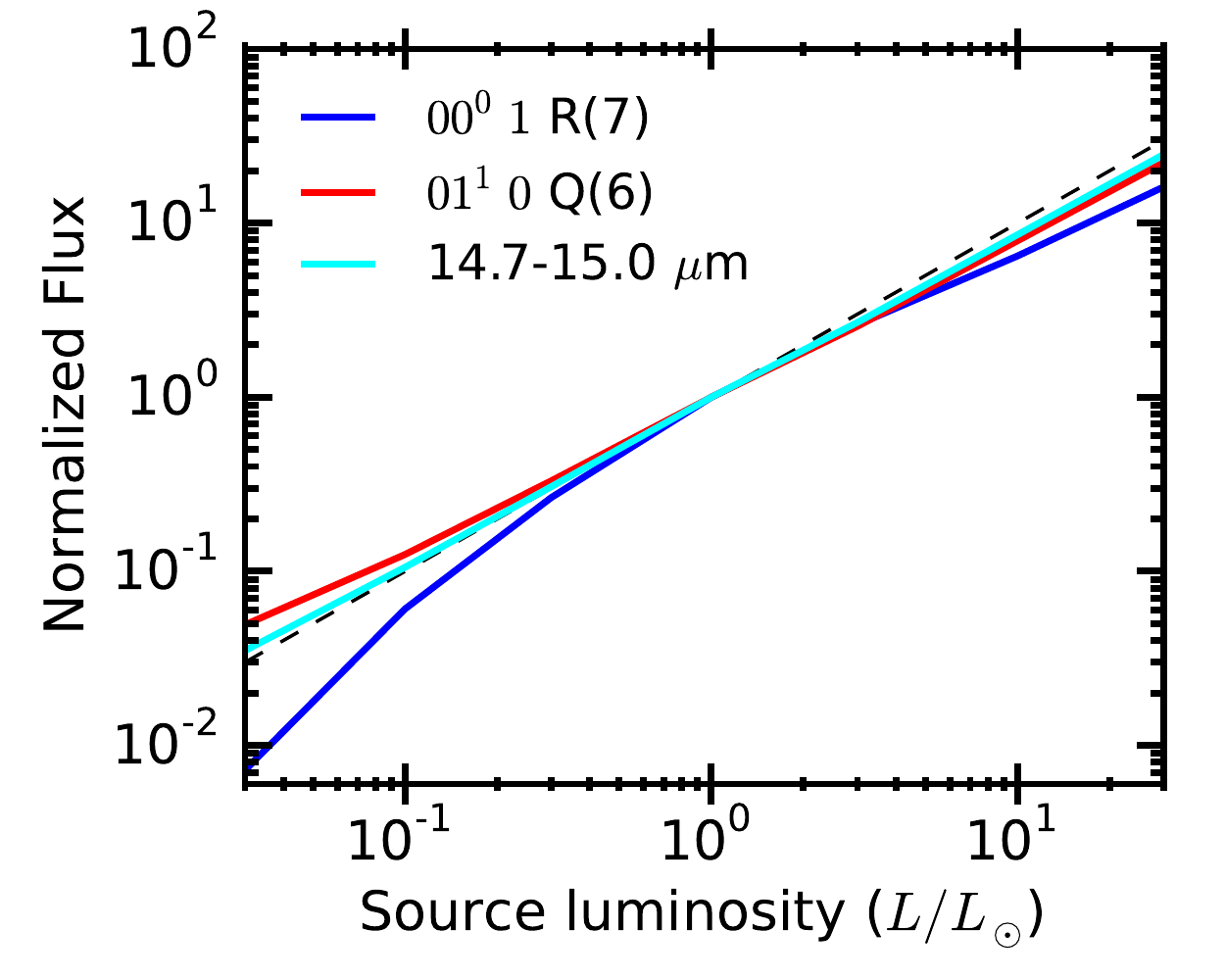}
\caption{\label{fig:lum_flux} \ce{CO2} flux versus stellar luminosity for three different features in the spectrum. All fluxes have been normalized to the flux of the 1 $L_{\odot}$ model. Both the flux in the 14.7 to 15 $\mu$m region, which corresponds to the $v_2$ $Q$-branch, and the flux from the $01^10$ $Q(6)$ line have a nearly linear dependence on the stellar luminosity. The dependence of the $00^01$ $R(7)$ line at 4.3 $\mu$m on source luminosity is more complex. }

\end{figure}

\begin{figure*}
\centering
\includegraphics[width=\hsize]{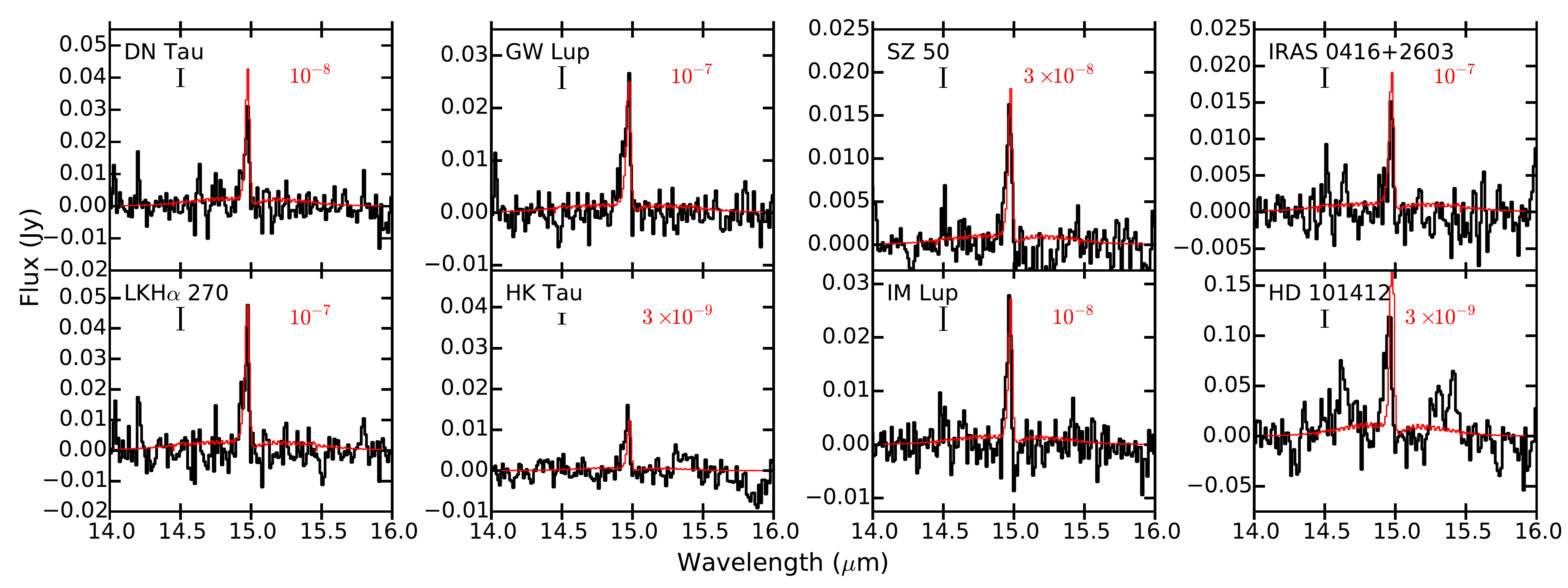}
\caption{\label{fig:cont_sub} Continuum subtracted spectra (black) with DALI {\ce{CO2}} emission models (red) for the eight selected sources. The abundance used in the DALI model is given in each frame, taken to be constant over the whole disk. The line fluxes have been corrected for the distances to the sources. The errorbar in the top left corner of each panel shows the median errorbar on the data.}
\end{figure*}

Two sets of comparisons are made. For the first set, the model fluxes
are only corrected for the distance to the objects. For the other set,
the model fluxes are scaled for the distance but also scaled for the
luminosity of the central source, using
$L_{\ce{CO2}} \propto L_\star$. This relation is found by running a
set of models with a range of luminosities, presented in
Fig.~\ref{fig:lum_flux}.  Aside from the luminosity of the star, all
other parameters have been kept the same including the shape of the
stellar spectrum. The effective temperature of the star mostly affects
the fraction of short wavelength UV photons which can photodissociate
molecules, but since no detailed chemistry is included, the use of a
different stellar temperature would not change our results.  Other
tests (not shown here) have indeed shown that the shape of the
spectrum does not really matter for the \ce{CO2} line fluxes in this
parametric model.  All models have a gas-to-dust ratio of 1000 and a
constant \ce{CO2} abundance of $10^{-8}$.

Both the total flux in the range between 14.7 and 15.0 $\mu$m and that
of a single line in this region (the $01^10$ $Q(6)$ line) have an
almost linear relation with luminosity of the central star. For the
$00^01$ $R(7)$ line around 4.3 $\mu$m the dependence on the central
luminosity is slightly more complex. Below a stellar luminosity of
$1\,L_\odot$ the dependence is stronger than linear, but above that the
dependence becomes weaker than linear.  Overall, it is reasonable to
correct the 15 $\mu$m fluxes from our model for source stellar
luminosities using the linear relationship. This is because the amount of infrared continuum radiation that the disk produces scales linearly with the amount of energy that is put into the disk by the stellar radiation. It is the infrared continuum radiation that sets the molecular emission the due to radiative pumping, the dominant vibrational excitation mechanism for \ce{CO2}.

The model spectra are overplotted on the continuum subtracted
observations in Fig.~\ref{fig:cont_sub}. The flux in these models has
been scaled with the distance of the source and the luminosity of the
central star.  A gas-to-dust ratio of 1000 is adopted as inferred from \ce{H2O} observations \citep{Meijerink2009}.

\begin{table*}
\centering
\caption{\label{tab:stellarparam} Stellar parameters.}
\begin{tabular}{l c c c c r}
\hline
\hline
Object & Source & Spectral& Stellar& Distance & References \\
& luminosity ($L_\odot$) & type & mass ($M_\odot$) & (pc) \\
\hline
DN Tau/Sz 82    & 1.62 & M0   &0.62  & 140  & \cite{Rigliaco2015}\\
GW Lup/Sz 71    & 0.3  & M1.5 & 0.42 & 150  & \cite{Alcala2014}\\
SZ50            & 0.57 & M4   & 0.27 & 178  & \cite{Manara2016}\\
IRAS 04216+2603 & 0.2    & M0   & 0.57 & 140 & \cite{Rigliaco2015}, \cite{Andrews2013}\\
LkH$\alpha$ 270 & 1.4 & K2 & ? & 235 & \cite{Winston2010}\\
HK Tau & 1.12 & M0.5 & 0.56 & 140 & \cite{Rigliaco2015} \\
IM Lup & 3.53  & M0 & 0.56 & 190 & \cite{Rigliaco2015} \\
HD101412 & 25 & A0 & 2.3 & 160 & \cite{vanderPlas2008} \\
\hline
\end{tabular}
\end{table*}

An overview of the inferred abundances is given in
Table~\ref{tab:infer}. For the DALI models the emitting \ce{CO2}
column and the number of emitting \ce{CO2} molecules have been tabulated in
Table~\ref{tab:infer}. The column is defined as the column of \ce{CO2} above the
$\tau_\mathrm{dust} =1$ line at the radial location of the peak of the contribution function (Fig.~\ref{fig:Gridplotoverview}d).
The number of molecules is taken over the region that is responsible for half the total emission as given by the contribution function. The number of molecules shown is thus the minimal amount of \ce{CO2} needed to explain the majority of the flux and a sets lower limit for the amount of \ce{CO2} needed to explain all of the emission. The inferred abundances range from $10^{-9} - 10^{-7}$.
They agree with that inferred by \cite{Pontoppidan2014} using an
LTE disk model appropriate for the RNO 90 disk, demonstrating that
non-LTE excitation effects are minor (see also Appendix~\ref{app:LTEvsNLTE}).

For GW Lup and SZ 50 the emitting \ce{CO2}  columns found by
\cite{Salyk2011} are within a factor of two from those inferred here, whereas for DN Tau and IM Lup our inferred columns are consistent with the upper limits from the slab models (tabulated in Table~\ref{tab:infer}).
The inferred column for HD101412 differs greatly however. For all disks the number of molecules in our models is at least an order of magnitude higher than the number of molecules inferred from the LTE models.
The emitting area used by \cite{Salyk2011} in fitting the
\ce{CO2} feature was fixed and generally taken to be slightly larger
than the inner 1 AU. This is very small compared to the emitting area
found in this work which extends up to 30 AU. It is thus not
surprising that the total number of \ce{CO2} molecules inferred is
lower for the LTE slab models from \cite{Salyk2011}. The high number of molecules needed for the emission in our models is also related to the difference in excitation: the
vibrational excitation temperature of the gas in the non-LTE models is
lower (100--300 K) than the temperature fitted for the LTE models
($\sim 650$ K). Thus in the non-LTE models a larger number of
molecules is needed to get the same total flux. The narrow \ce{CO2} profile is due to low rotational temperatures as emission from large radii > 2--10 AU dominates the strongest lines. The visual contrast is enhanced by the fact that the \ce{CO2} feature is also intrinsically narrower at similar temperatures than that of \ce{HCN} (Fig.~\ref{fig:CO2_HCN}). For HD 101412 the model feature is
notably narrower than the observed feature signifying either a
higher {\ce{CO2}} rotational temperature, or a more optically thick
emitting region.

There are of course caveats in the comparisons done here. The standard
model uses a T-Tauri star that is luminous (total luminosity of
$7.3\, L_\odot$) and that disk is known to have very strong \ce{H2O}
emission. The sample of comparison protostars consists of 7 T-Tauri
stars with luminosities a factor of 2 -- 35 lower and a Herbig Ae
star that is more than 3 times as luminous. A simple correction for
source luminosity is only an approximation. All of these
sources have little to no emission lines of \ce{H2O} in the
mid-infrared. This may be an indication of different disk structures, and the
disk model used in this work may not be representative of these
water-rich disks.  Indeed, \cite{Banzatti2015} found
that the emitting radius of the CO ro-vibrational lines scales
inversely with the vibrational temperature inferred from the CO
emission. This relation is consistent with inside-out gap opening.
Comparing the CO ro-vibrational data with \ce{H2O} infrared emission data
from VLT-CRIRES and Spitzer-IRS \cite{Banzatti2016} found that
there is a correlation between the radius of the CO ro-vibrational
emission and the strength of the water emission lines: The larger the
radius of the CO emission, the weaker the \ce{H2O} emission. This suggests
that the \ce{H2O}-poor sources may also have inner gaps, where both CO and
\ce{H2O} are depleted. There are only two \ce{H2O}-poor sources in our sample that overlap with \cite{Banzatti2016}.

However, if our analysis is applied to sources that do have water
emission, the range of best-fit \ce{CO2} abundances is found to be
similar. Fig. G.2 shows \ce{CO2} model spectra compared to observations for
a set of the strongest water emitting sources. The conclusion that the
abundance of \ce{CO2} in protoplanetary disks is around $10^{-8}$ is therefore
robust.

\begin{table*}
\centering
\caption{\label{tab:infer} Inferred \ce{CO2} abundances from Spitzer data}
\begin{tabular}{l c c c c c c}
\hline
\hline
\multicolumn{1}{l}{Object} & \multicolumn{2}{c}{Inferred abundance (w.r.t \ce{H})}  & \multicolumn{2}{c}{\cite{Salyk2011} LTE slab results}& \multicolumn{2}{c}{DALI Non-LTE results} \\
 &  $d$ corr. & $d$ and $L_\star$ corr.  & Column (cm$^{-2}$) & $\mathcal{N}_{\mathrm{tot},{\ce{CO2}}}$ (mol.) & Column (cm$^{-2}$) & $\mathcal{N}_{\mathrm{tot},\ce{CO2}}$ (mol.)\tablefootmark{a} \\
\hline
DN Tau   		& $1(-9)$ 		& $1(-8)$	& $<5.0(14)$	& $<3.8(41)$	&2(14)		&2(44)	\\
GW Lup   		& $1(-9)$ 		& $1(-7)$	& $1.5(15)$		& $1.1(42)$		&2(15)		&8(44)	\\
SZ50            & $1(-9)$ 		& $3(-8)$ 	& $1.0(15)$		& $7.6(41)$		&7(14)		&5(44)	\\
IRAS 04216+2603 & $<1(-9)$ 		& $1(-7)$ 	& --			& --			&2(15)		&8(44)	\\
LkH$\alpha$ 270	& $0.3 - 1(-8)$ & $1(-7)$	& $3.9(15)$		& $4.1(42)$		&2(15)		&8(44)	\\
HK Tau 			& $<1(-9)$ 		& $3(-9)$	& -- 			& --			&5(13)		&1(44)	\\
IM Lup 			& $1-3(-9)$ 	& $1(-8)$	& $<7.9(14)$	& $<6.9(41)$	&2(14)		&2(44)	\\
HD 101412 		& $1(-8)$ 		& $1-3(-9)$	& $1.0(16) $	& $1.94(43)$	&$2-5(13)$	&0.7--1(44)	\\
\hline
\end{tabular}
\tablefoot{$a(b) = a \times 10^{b}$\\
\tablefoottext{a}\ Minimal number of \ce{CO2} molecules responsible for half of the emission 
}
\end{table*}

The inferred \ce{CO2} abundances are low, much lower
than the expected ISM value of $10^{-5}$ if all
of the \ce{CO2} would result from sublimated ices. This demonstrates that
the abundances have been reset by high temperature chemistry, as also concluded by \citet{Pontoppidan2014}.

The inferred low abundances agree well with chemical models by
\cite{Walsh2015}. However, the column found for chemical models by \cite{Agundez2008},
$\sim 6\times 10^{16}$ cm$^{-2}$, is more than an order of magnitude
higher. \cite{Agundez2008} used a different lower vertical bound for
their column integration and only considered the inner 3 AU. Either of
these assumptions may explain the difference in the \ce{CO2} column.

\subsection{Tracing the {\ce{CO2}} iceline}
\label{ssc:Iceline}
One of the new big paradigms in (giant) planet formation is pebble
accretion. Pebbles, in models defined as dust particles with a Stokes
number around 1, are badly coupled to the gas, but generally not
massive enough to ignore the interaction with the gas. This means that
these particles settle to the mid-plane and radially drift inward on
short time scales. 
This pebble flux allows in theory a planetesimal to accrete all the
pebbles that form at radii larger than its current location
\citep{Ormel2010,Lambrechts2012,Levison2015}.

This flux of pebbles also has consequences for the chemical
composition of the disk. These pebbles should at some point encounter
an iceline, if they are not stopped before. At the iceline they
should release the corresponding volatiles. The same holds for any
drifting planetesimals \citep{Ciesla2006}. As the ice composition is
very different from the gas composition, this can in principle
strongly change the gas content in a narrow region around the ice
line. For this effect to become observable in mid-infrared lines, the
sublimated ices should also be mixed vertically to higher regions in
the disk.

From chemical models the total gas-phase abundance of {\ce{CO2}}
around the {\ce{CO2}} iceline is thought to be relatively low, \citep[$10^{-8}$,][]{Walsh2015}
similar to the value found in this work. The {\ce{CO2}}
ice content in the outer disk can be orders of magnitude higher. Both
chemical models and measurements of comets show that the {\ce{CO2}}
content in ices can be more than 20\% of the total ice content
\citep{LeRoy2015,Eistrup2016}, with \ce{CO2} ice even becoming more
abundant than \ce{H2O} ice in some models of outer disk chemistry
\citep{Drozdovskaya2016}. This translates into an abundance up to a few $\times\,10^{-5}$. We investigate here
whether the evaporation of these {\ce{CO2}} ices around the iceline
would be observable.

\begin{figure*}
\includegraphics[width=\hsize]{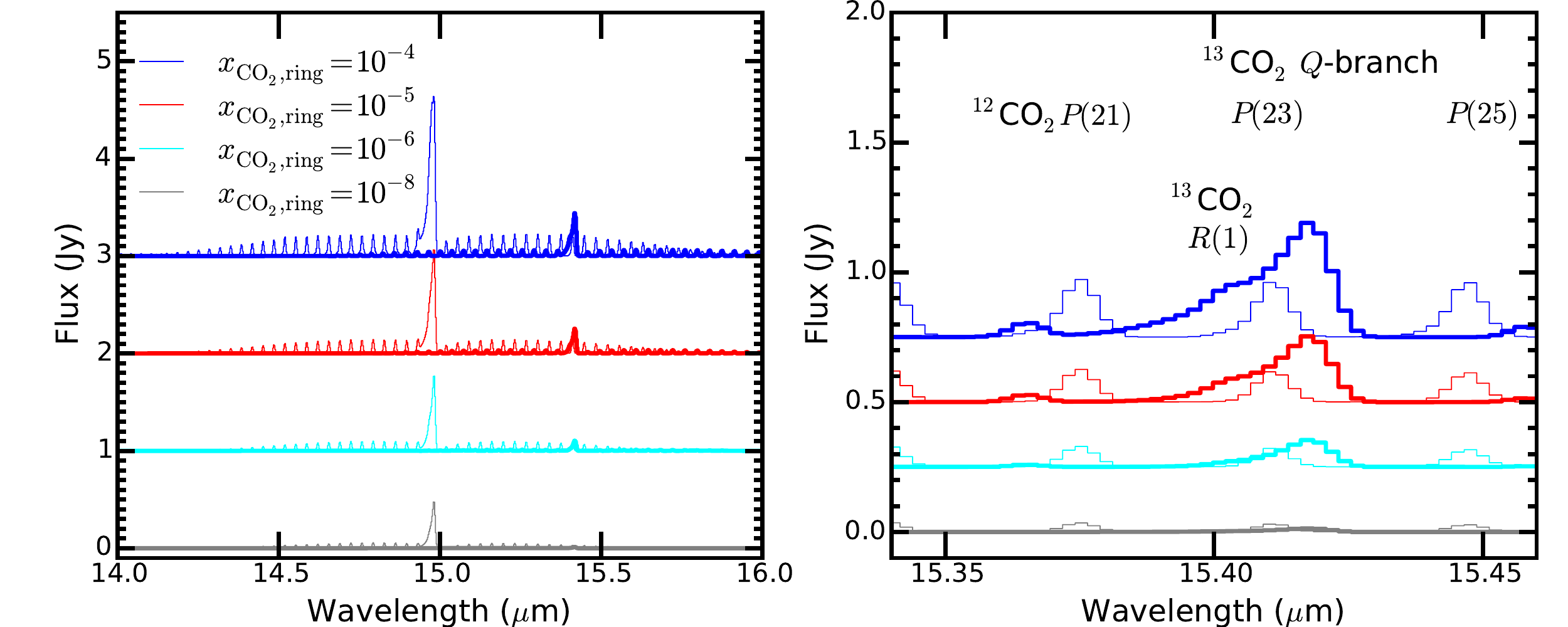}
\caption{\label{fig:ringspectra} \ce{^{12}CO2} (thin lines) and \ce{^{13}CO2} spectra (thick lines) of models with an enhanced {\ce{CO2}} abundance around the \ce{CO2} iceline. The models have a constant abundance of {\ce{CO2}} of $10^{-8}$ throughout most of the disk. The grey lines denote the model without any local enhancements in the {\ce{CO2}} abundance in an annulus around the \ce{CO2} iceline. The cyan, red and blue lines show the models with an enhanced abundance of $10^{-6}$, $10^{-5}$ and $10^{-4}$, respectively. The \ce{^{12}CO2} flux is enhanced by a factor of 2 to 4 over the complete range of the spectra. The \ce{^{13}CO2} $Q-$branch however becomes more than an order of magnitude stronger, and the peak flux of the feature becomes higher than the peak flux of the neighbouring \ce{^{12}CO2} lines for the highest abundances. The spectra with enhanced {\ce{CO2}} abundances are shifted vertically for clarity.}
\end{figure*}

To model the effect of pebbles moving over the iceline, a model with
a constant {\ce{CO2}} abundance of $1 \times 10^{-8}$ and a
gas-to-dust ratio of 1000 is taken. In addition, the abundance of
{\ce{CO2}} is enhanced in an annulus where the midplane temperature is
between 70 K and 100 K (grey region in Fig.~\ref{fig:Gridplotoverview}b), corresponding to the sublimation temperature of
pure \ce{CO2} ice \citep{Harsono2015}. This results in a radial
area between 8 and 15 AU in our case. The abundance is taken to be
enhanced over the total vertical extent of the {\ce{CO2}} in the
model, as in the case of strong vertical mixing. The spectra from
three models with enhanced abundances of
$x_{\ce{CO2},\mathrm{ring}} = 10^{-6}, 10^{-5}$ and $10^{-4}$ in this
ring can be seen in Fig.~\ref{fig:ringspectra} together with the
spectrum for the constant $x_{\ce{CO2}} = 10^{-8}$ model. 
Note that \ce{CO2} ice is unlikely to be pure, and that some of it
will likely also come off at the \ce{H2O} iceline, but such a
multi-step sublimation model is not considered here.

The spectra in Fig.~\ref{fig:ringspectra} show both \ce{^{12}CO2} and
\ce{^{13}CO2} in thin and thick lines respectively for the model with
a constant abundance (grey), an enhanced abundance of
$1\times 10^{-6}$ (cyan), $1\times 10^{-5}$ (red) and
$1\times 10^{-4}$ (blue) around the \ce{CO2} iceline.  The enhanced
{\ce{CO2}} increases the \ce{^{12}CO2} flux up to a factor of 3. The
optically thin \ce{^{13}CO2} feature is however increased by a lot
more, with the peak flux in the \ce{^{13}CO2} $Q-$branch reaching
fluxes that are more than two times higher than peak fluxes of the nearby
\ce{^{12}CO2} $P-$branch lines.

Enhanced abundances in the outer regions can be distinguished from an
enhanced abundance in the inner regions by looking at the tail of the
\ce{^{13}CO2} $Q-$branch. An over-abundant inner region will show a
significant flux (10--50\% of the peak) from the \ce{^{13}CO2}
$Q-$branch in the entire region between the locations of the
\ce{^{12}CO2} $P(21)$ and $P(23)$ lines and will show a smoothly
declining profile with decreasing wavelength. If the abundance
enhancement is in the outer regions, where the gas temperature and
thus the rotational temperature is lower, the flux between the
\ce{^{12}CO2} $P(21)$ and $P(23)$ lines will be lower for the low
enhancements (0--20\% of the peak flux), for the higher enhancements
the $R$(1) line shows up in the short wavelength side of the profile. Other $R$ and $P$ branch lines from
\ce{^{13}CO2} can also show up in the spectrum if the abundance can
reach up to $10^{-4}$ in the ring around the \ce{CO2} iceline.

\subsection{Comparison of {\ce{CO2}} with other inner disk molecules}

With the models presented in this paper, four molecules with
rovibrational transitions coming from the inner disk have been studied
by non-LTE disk models: \ce{H2O}
\citep{Antonelli2015,Antonelli2016,Meijerink2009}, \ce{CO}
\citep{Thi2013}, \ce{HCN} \citep{Bruderer2015} and {\ce{CO2}} (this
work). Of these \ce{CO} is special, as it can be excited by UV
radiation and fluoresces to excited vibrational states that in turn
emit infrared radiation. For the other molecules absorption of a UV
photon mostly leads to dissociation of the molecule \citep{Heays2017}. \ce{H2O}, \ce{HCN} and \ce{CO} all have an permanent dipole
moment and can thus also emit strongly in the sub-millimeter. This means that
these molecules will have lower rotational temperatures than
{\ce{CO2}} in low density gas but they are actually observed to have broader profiles in the mid-infrared. Thus, our models reinforce the conclusion from the observed profiles that \ce{CO2} comes from relatively cold gas (200-300 K) (See panel \textit{c} of Fig.~\ref{fig:Gridplotoverview} and panel \textit{d} of Fig.~\ref{fig:Overview_app}).

For the disk around AS 205 (N) the emission of both \ce{HCN} and \ce{CO2} has now been analysed under non-LTE conditions in DALI.
 As such it is possible to infer the \ce{HCN} to \ce{CO2} abundance ratio in the disk. The representative, constant abundance model for the \ce{CO2} emission from AS 205 N has an abundance of $3\times 10^{-8}$ with a gas-to-dust ratio of 1000 but the inner abundance can easily vary by an order of magnitude or more while still being in agreement with observations. The models from \cite{Bruderer2015} that best reproduce the data have outer \ce{HCN} abundances between $10^{-10}$ and $10^{-9}$ for gas-to-dust ratios of 1000.
 This translates into \ce{CO2}/\ce{HCN} abundance ratios of 30--300 in the region from 2 to 30 AU. The higher abundance of \ce{CO2} in the outer regions of the disk explains the colder inferred rotational temperature of \ce{CO2} compared to \ce{HCN}.

\section{Conclusion}

Results of DALI models are presented, modelling the full continuum
radiative transfer, non-LTE excitation of {\ce{CO2}} in a typical
protoplanetary disk model, with as the main goal: to find a way to
measure the \ce{CO2} abundance in the emitting region of disks with
future instruments like JWST and test different assumptions on its origin.  Spectra of \ce{CO2} and \ce{^{13}CO2}
in the 4--4.5 $\mu$m and 14--16 $\mu$m regions were modelled for disks
with different parametrized abundance structures, gas masses and dust
masses. The main conclusions of this study are:

\begin{itemize}

\item The critical density of the {\ce{CO2}} $00^01$ state, responsible for emission around 4.3 $\mu$m, is very high, $>10^{15}$ cm$^{-3}$. As a result, in the absence of a pumping radiation field, there is no emission from the $00^01$ state at low densities. If there is a pumping infrared radiation field, or if the density is high enough, the emission around 4.3 $\mu$m will be brighter than that around 15 $\mu$m. 

\item The infrared continuum radiation excites {\ce{CO2}} up to large
  radii (10s of AU). The region probed by the {\ce{CO2}} emission
  can therefore be an order of magnitude larger (in radius) than typically assumed in
  LTE slab models. Temperatures inferred from optically thin LTE models can also be larger than the actual temperature of the emitting gas. Differences between LTE and non-LTE full disk model flux are typically a within factor of three.

\item Current observations of the 15 $\mu$m $Q-$branch fluxes
  are consistent with models with constant abundances between $10^{-9}$ and
  $10^{-7}$ for a gas-to-dust ratio of 1000. Observations of lines corresponding to levels
  with high rotational quantum numbers or the \ce{^{13}CO2} $Q-$branch
  will have to be used to properly infer abundances. Especially
  the \ce{^{13}CO2} $Q-$branch can be a good indicator of abundance
  structure from inner to outer disk.

\item The gas-to-dust ratio and fractional abundance are largely
  degenerate. What sets the emission is the column of {\ce{CO2}} above
  the dust infrared photosphere. Models with similar columns have very
  similar spectra irrespective of total dust and gas mass, due to the excitation mechanism of \ce{CO2}. 
   If the gas-to-dust ratio is constrained from other observations such as \ce{H2O} the fractional abundance can be determined from the spectra. 

\item The abundance of \ce{CO2} in protoplanetary disks inferred from modelling, $10^{-9}$--$10^{-7}$, is at least 2 orders of magnitude lower than the \ce{CO2} abundance in ISM ices. This implies that disk chemical abundances are not directly inherited from the ISM and that significant chemical processing happens between the giant molecular cloud stage and the protoplanetary disk stage.  

\item The \ce{^{13}CO2} $v_2$ Q-branch at 15.42 $\mu$m will be able to
  identify an overabundance of {\ce{CO2}} in the upper layers of the inner disk, such as could be produced by sublimating pebbles and planetesimals around the iceline(s).

\end{itemize} 

Our work shows that the new instruments on JWST will be able to give a wealth of information on the \ce{CO2} abundance structure, provided that high $S/N$ ($>300$ on the continuum) spectra are obtained. 

\section*{Acknowledgements}
We thank the anonymous referee for his/her suggestions that have improved the paper. Astrochemistry in Leiden is supported by the European Union A-ERC grant 291141 CHEMPLAN, by the Netherlands Research School for Astronomy (NOVA), by a Royal Netherlands Academy of Arts and Sciences (KNAW) professor prize. This work is based in part on archival data obtained made with the Spitzer Space Telescope, which is operated by the Jet Propulsion Laboratory, California Institute of Technology under a contract with NASA.
\bibliographystyle{aa}
\bibliography{Lit_list}
\begin{appendix}
\section{Collisional rate coefficients}
\label{app:rovib}
The collisional rate coefficients are calculated in a way very similar to \cite{Bruderer2015}, that is by combining vibrational coefficients with rotational rate coefficients to get the state-to-state ro-vibrational rate coefficients. Only collisions with \ce{H2} are considered, which is the dominant gas species in the regions were \ce{CO2} is expected to be abundant. The vibrational coefficients were taken from the laser physics and atmosperic physics papers. An overview of the final vibrational rate coefficients used are shown in Table~\ref{tab:vibratecoef}. The temperature dependence of the rates is suppressed for the de-excitation collissional rate coefficients and the rate for 300 K are used throughout. The vibrational rate coefficients are not expected to vary much over the range of temperatures considered here. The de-excitation rate coefficient of the bending mode by \ce{H2} ($01^10 \rightarrow 00^00$ transition) from \cite{Allen1980} is $5\times 10^{-12}$ cm$^{-3}$s$^{-1}$, an order of magnitude faster than the \ce{He} \citep{Taylor1969surveyCO2,Allen1980}. This is probably due to vibrational-rotational energy exchange in collisions with rotationally excited \ce{H2} \citep{Allen1980}. For levels higher up in the vibrational ladder we extrapolate the rates as done by \cite{Proccaccia} and \cite{Chandra2001}. Combining Eq. 6 and 8 from the last paper we get, for $v > w$, the relation:
\begin{equation}
k(v_2 = v \rightarrow v_2 = w) = v\frac{2w+1}{2v+1} k(v_2 = 1 \rightarrow v_2 = 0)\textrm{.}
\end{equation}

The rate coefficient measured by \cite{Nevdakh2003} is actually the total quenching rate of the $00^01$ level. Here we assume that the relaxiation of the $00^01$ level goes to the three closest lower energy levels ($03^30, 11^10(1), 11^10(2)$) in equal measure. For the all the rates between the levels of the Fermi degenerate states and the corresponding bending mode with higher angular momentum the \ce{CO2-CO2} rate measured by \cite{Jacobs1975_0200_relax} was used scaled to the reduced mass of the \ce{H2}-{\ce{CO2}} system. The states with constant $2 v_1 + v_2$ are considered equal to the pure bending mode with respect to the collisional rate coefficients to other levels.

\begin{table}
\centering
\caption{\label{tab:vibratecoef} Vibrational rate coefficients as measured/extrapolated.}
\begin{tabular}{l c r}
\hline \hline
Initial state & Final state & Rate coef. [cm$^3$s$^{-1}$]\\
\hline
$01^10$ & $00^00$& 5(-12)\\
$10^00(2)$ & $00^00$& 3.3(-12)\\
$10^00(2)$ & $01^10$& 1(-11)\\
$02^20$ & $00^00$& 3.3(-12)\\
$02^20$ & $01^10$& 1(-11)\\
$02^20$ & $10^00(2)$& 5.8(-12)\\
$10^00(1)$ & $00^00$& 3.3(-12)\\
$10^00(1)$ & $01^10$& 1(-11)\\
$10^00(1)$ & $10^00(2)$,$02^20$& 5.8(-12)\\
$11^10(2)$ & $00^00(1)$& 3(-12)\\
$11^10(2)$ & $01^10(1)$& 9(-12)\\
$11^10(2)$ & $10^00(1,2)$,$02^20$ & 1.5(-11) \\
$03^30$ & $00^00$& 3(-12)\\
$03^30$ & $01^10$& 9(-12)\\
$03^30$ & $10^00(1,2)$,$02^20$ & 1.5(-11) \\
$03^30$ & $11^10(2)$& 5.8(-12) \\
$11^10(1)$ & $00^00$& 3(-12)\\
$11^10(1)$ & $01^10$& 9(-12)\\
$11^10(1)$ & $10^00(1,2)$,$02^20$ & 1.5(-11) \\
$11^10(1)$ & $11^10(2), 03^30$& 5.8(-12) \\
$00^01$ & $11^10(1,2), 03^30$ & 4.5(-14) \\
$01^11$ & $11^10(1,2), 03^30$ & 4.5(-14) \\
$01^11$ & $00^01$ & 5(-12) \\
\hline\\

\end{tabular}
\tablefoot{$a(b) = a \times 10^{b}$}

\end{table}

No information exists on the rotational rate coefficients of {\ce{CO2}} with \ce{H2}. We have decided to use the \ce{CO} rate coefficients from \cite{Yang2010} instead. Since {\ce{CO2}} does not have a dipole moment, the exact rate coefficients are not expected to be important since the critical densities for the levels in the rotational ladders are very low, $ < 10^{4}$ cm$^{-3}$.

The method suggested by \cite{Faure2008} was employed to calculate the state-to-state de-excitation rate coefficients for initial levels $v,J$ to all levels $v',J'$ with a smaller ro-vibrational energy. This is assuming a decoupling of rotational and vibrational levels, so we can write:

\begin{equation}
k(v,J \rightarrow v',J';T) = k(v \rightarrow v') \times P_{(J,J')}(T),
\end{equation}
where 
\begin{equation}
P_{(J,J')}(T) = \frac{k(0,J\rightarrow 0,J';T)\sum_J g_J \exp(-E_{v,J}/kT) }{\sum_J g_J \exp(-E_{v,J}/kT) \sum_{J'} k(0,J\rightarrow 0,J';T)},
\end{equation}
with the statistical weights $g_i$ of the levels. All the excitation rates are calculated using the detailed balance. 

\section{Fast line ray tracer}
\label{app:fastray}
For the calculation of the \ce{CO2} lines a new ray tracer was used. The conventional ray tracer used in DALI \citep{Bruderer2012} can take up to 10 minutes to calculate the flux from one line. The \ce{CO2} molecule model used here includes more than 3600 lines. Not al of these lines are directly important but to get the complete spectrum of both the 4.3 $\mu$m and the 15 $\mu$m bands, a few hundred lines need to be ray traced for each model.

To enable the calculation of a large number of lines a module has been implemented into DALI that can calculate a line flux in a few seconds versus a few minutes for the conventional ray tracer. The module uses the fact that, along a line of sight, the velocity shear due to the finite height of the disk is approximately linear \citep[][Eqs. 9 and 10]{Horne1986}. Using this, the spectrum for an annulus in the disk can be approximated. At the radius of the annulus in question, the spectra are calculated for different velocities shears. These spectra are calculated by vertically integrating the equation of radiative transfer through the disk and correcting for the projected area for non face-on viewing angles. Then the total spectrum of the annulus is calculated by iterating over the azimuthal direction. For each angle the velocity shear is calculated and the spectrum is interpolated from the pre-calculated spectra. A simple sum over the spectra in all annuli is now sufficient to calculate the total spectrum.

This approximation is a powerful tool for calculating the total flux in a line especially for low inclinations. For the models presented in this paper the fluxes differ by about 4\% for the 15 $\mu$m lines and 1.5\% for the 4.3 $\mu$m lines. This is small compared to the other uncertainties in the models.

The approximation breaks down at high inclinations and should be used with care for any inclination larger than 45$^\circ$. The total line shape is also close to the line shape from the traditional ray tracer, but with the high S/N from ALMA, using the traditional ray tracer is still advised for doing direct comparisons. This same goes for images for which the errors will be larger than for the integrated flux or line shape as some of the errors made in making the image will cancel out (in first order) when integrating over the annulus.

\section{Model temperature and radiation structure}
\begin{figure*}
\includegraphics[width=\hsize]{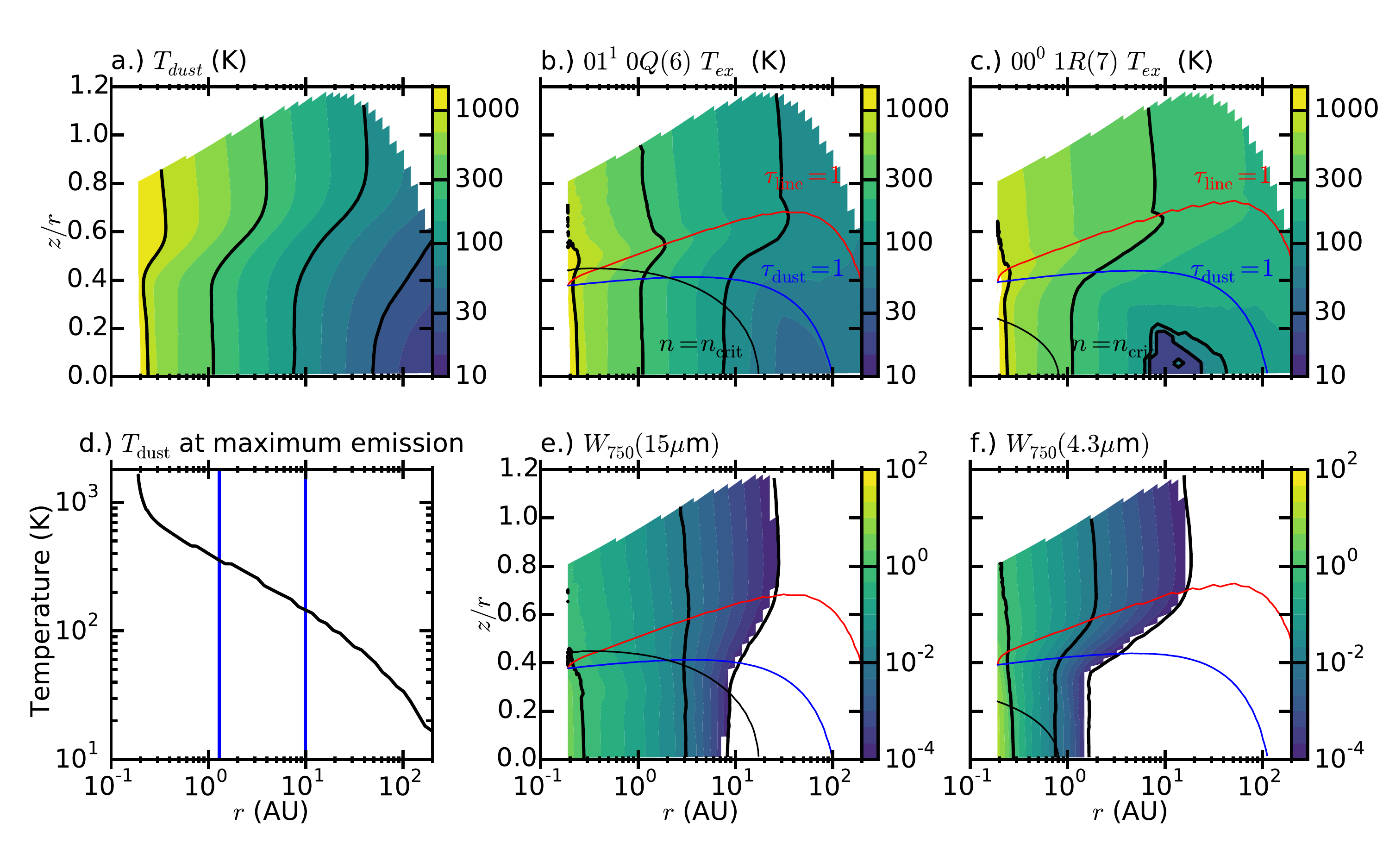}
\caption{\label{fig:Overview_app} Dust temperature, excitation temperature and radiation field for a model with $g/d_\mathrm{gas}=1000$, and a constant abundance of $10^{-7}$. The red line in panels \textit{b},\textit{c},\textit{e} and \textit{f} shows the \ce{CO2} line $\tau = 1$ surface, the blue line shows the $\tau = 1$ line for the dust. Panel \textit{d} shows the dust temperature at the height from which most of the emission of the 15 $\mu$m $Q(6)$ line originates as function of radius. The vertical blue lines enclose the radii that account for 50\% of the emission.} 
\end{figure*}

Fig.~\ref{fig:Overview_app} shows the model temperature, radiation field and excitation temperature structure corresponding to the model shown in Fig.~\ref{fig:Gridplotoverview}. Panel \textit{a} shows the dust temperature structure, panels \textit{b} and \textit{c} show the excitation temperature of the $v_2$ $1\rightarrow 0$ $Q(6)$ line and $v_3 $ $1\rightarrow 0$ $R(7)$ lines. For the excitation temperature only the upper and lower state of the line are used. This is thus a vibrational excitation temperature and can be different from the ground state rotational excitation temperature (that follows the dust temperature) and the rotational excitation temperature within a vibrationally excited state. Where the density is higher than the critical density the excitation temperature is equal to the dust temperature. In the disk atmosphere the $Q(6)$ is mostly subthermally excited, while the R(7) line is superthermally excited. For both lines there is a maximum in the vertical excitation temperature distribution at the point where the gas becomes optically thick to its own radiation. Panel \textit{d} shows the dust temperature of the region from which most of the \ce{CO2} 15 $\mu$m emission originates as function of radius. Most of the emitting gas is at temperatures between 150 and 350 K. Panel \textit{e} and \textit{f} show the strength of the radiation field at 15 $\mu$m and 4.3 $\mu$m is shown as a factor of the radiation field of a 750 K blackbody. This shows where there is a sufficient photon density to radiatively pump {\ce{CO2}}.

\section{Model fluxes $g/d_{\mathrm{dust}}$}
\label{app:gd_dust}
\begin{figure}
\centering
\includegraphics[width=\hsize]{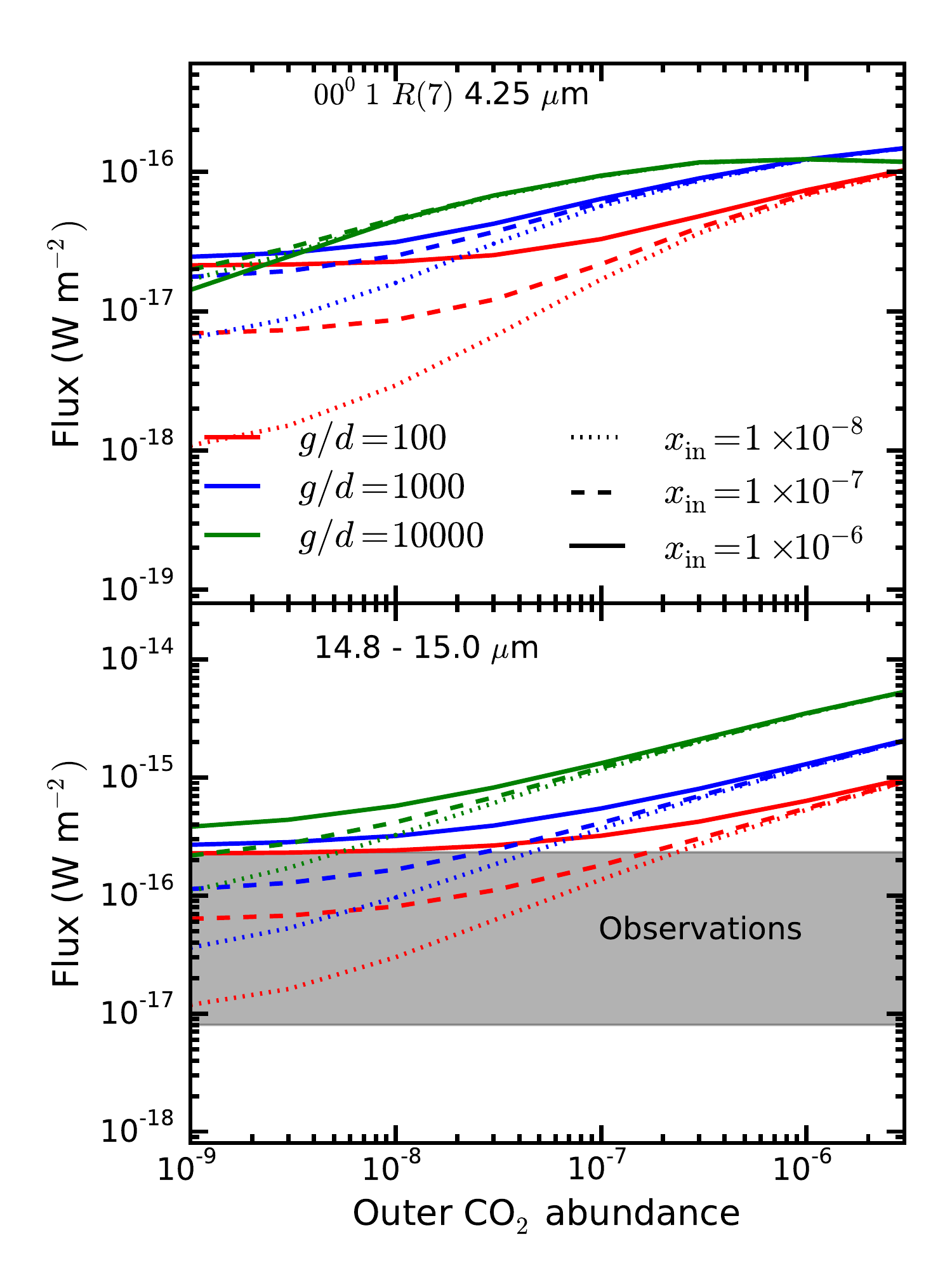}
\caption{\label{fig:Flux_dust} Line fluxes for models with constant gas mass ($g/d_{dust}$). The upper panel shows the flux of the R(7) line from the fundamental asymmetric stretch band at 4.3 $\mu$m. The lower panel shows the flux contained in the 15 $\mu$m feature. The grey region denotes the range in line fluxes as observed by \cite{Salyk2011} scale to the distance of AS 205 N. This feature contains the flux from multiple $Q-$branches with $\Delta v_2 =1 $. The {\ce{CO2}} flux depends primarily on the outer {\ce{CO2}} abundance and the total g/d ratio and does not strongly depend on the inner {\ce{CO2}} abundances. Only for the very low {\ce{CO2}} absolute abundances in the outer regions is the effect of the inner abundance on the line fluxes visible.}

\end{figure}

As mentioned in the main text, two different way of changing the gas-to-dust ratio were considered, increasing the gas mass and decreasing the dust mass w.r.t. the gas-to-dust ratio 100 case. Fig.~\ref{fig:Flux_dust} is the counterpart to Fig.~\ref{fig:Flux_gas} showing the modelled fluxes for different inner \ce{CO2} abundances, outer \ce{CO2} abundances and different gas-to-dust ratios. In this case the gas-to-dust ratio is varied by keeping the gas mass of the disk constant and decreasing the amount of dust in the disk.

There are only very slight differences between Fig.~\ref{fig:Flux_dust} and Fig.~\ref{fig:Flux_gas} and all observations made for the figure in the main text hold for this figure as well. 

\section{LTE vs non-LTE}
\label{app:LTEvsNLTE}
\begin{figure}
\centering
\includegraphics[width=\hsize]{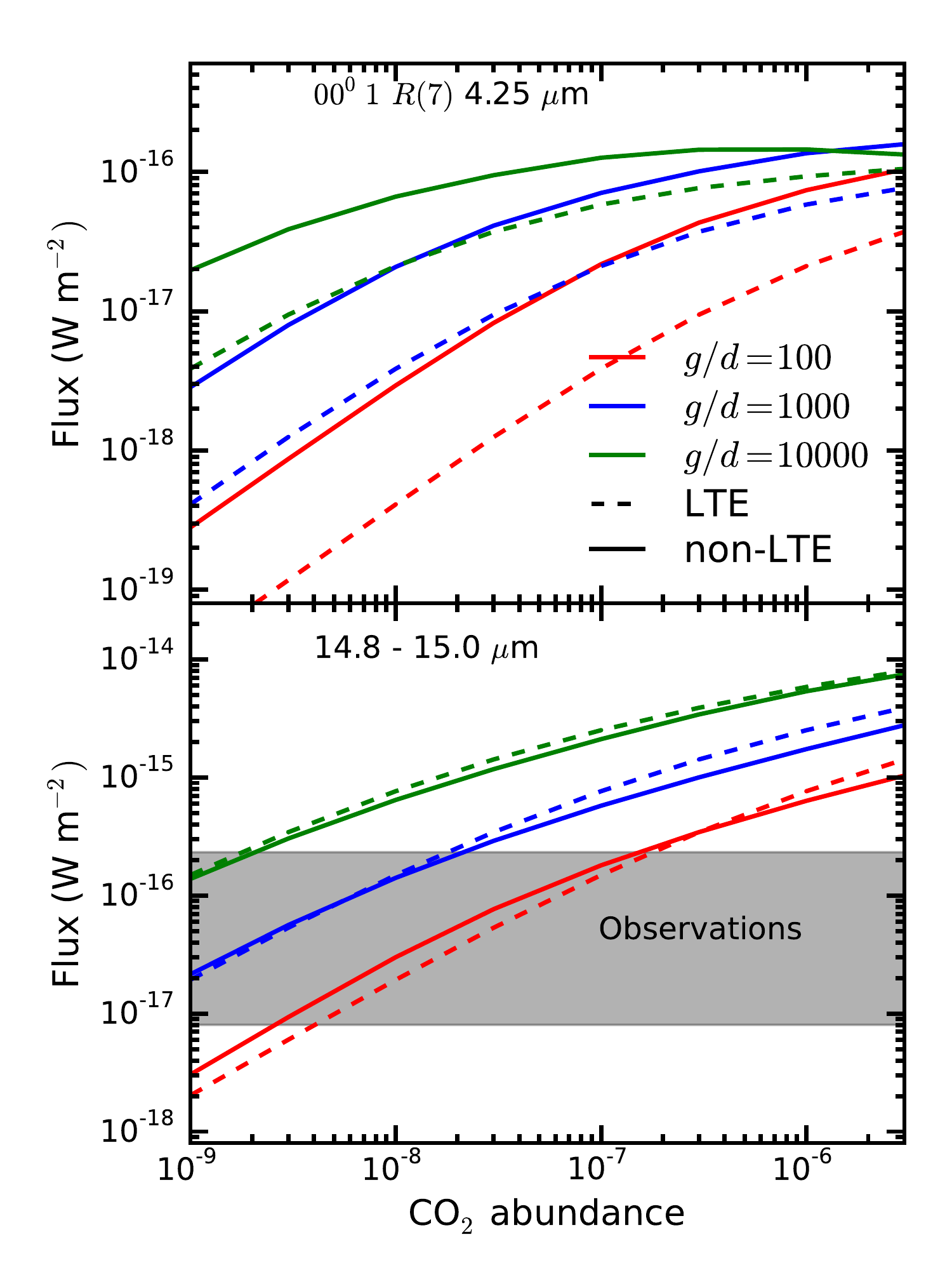}
\caption{\label{fig:flux_LTE} Flux comparison between LTE and non-LTE models. The upper panels show the flux of the R(7) line from the fundamental asymmetric stretch band. The lower panels show the flux contained in the 15 $\mu$m feature. The grey region shows the range of observed flux by \cite{Salyk2011}. The abundance in these models is constant over the whole disk. For the 15 $\mu$m feature the flux differences are small of the order of 30\%. The differences are more pronounced in the 4.25 $\mu$m flux where the differences can get as large as a factor of 4. }
\end{figure}

The effects of the LTE assumption on the line fluxes in a full disk model on the $v_3, 1\rightarrow 0, R(7)$ line and the 15 $\mu$m feature are shown in Fig.~\ref{fig:flux_LTE}. Only the models with constant abundance ($x_\mathrm{out}=x_\mathrm{in}$) are shown for clarity but the differences between LTE and non-LTE for these models are representative for the complete set of models. For the 15 $\mu$m flux the differences between the LTE and non-LTE models is small, of the order of 30\%. The radial extent of the emission is, however, different: the region emitting 75\% of the 15 $\mu$m flux extends twice as far in the non-LTE models (extent of the 15 $\mu$m non-LTE emission is seen in Fig.~\ref{fig:Gridplotoverview}, panel f). This a clear sign of the importance of infrared pumping that is included in the non-LTE models. The difference between the fluxes in the 4.25 $\mu$m line are greater, up to an order of magnitude. The difference are strongest in the models that have a low total {\ce{CO2}} content (so low abundance and low gas-to-dust ratio). This is mostly due to the larger radial extent of the emitting region extending up to 20 times further out in the non-LTE models compared to the corresponding LTE model (Extent of the 4.3 $\mu$m non-LTE emission is seen in Fig.~\ref{fig:Gridplotoverview}, panel i). This is in line with the higher Einstein $A$ coefficient and upper level energy (and thus higher critical density) of the $v_3, 1\rightarrow 0, R(7)$ line, giving rise to a large importance of infrared pumping relative to collisional excitation. Fig.~\ref{fig:flux_LTE} uses $g/d_{\mathrm{dust}} =1000$, but the plot for $g/d_{\mathrm{gas}} =1000$ is very similar.

\section{Line blending by \ce{H2O} and \ce{OH}}
\label{app:water}

\begin{figure}
\centering
\includegraphics[width=\hsize]{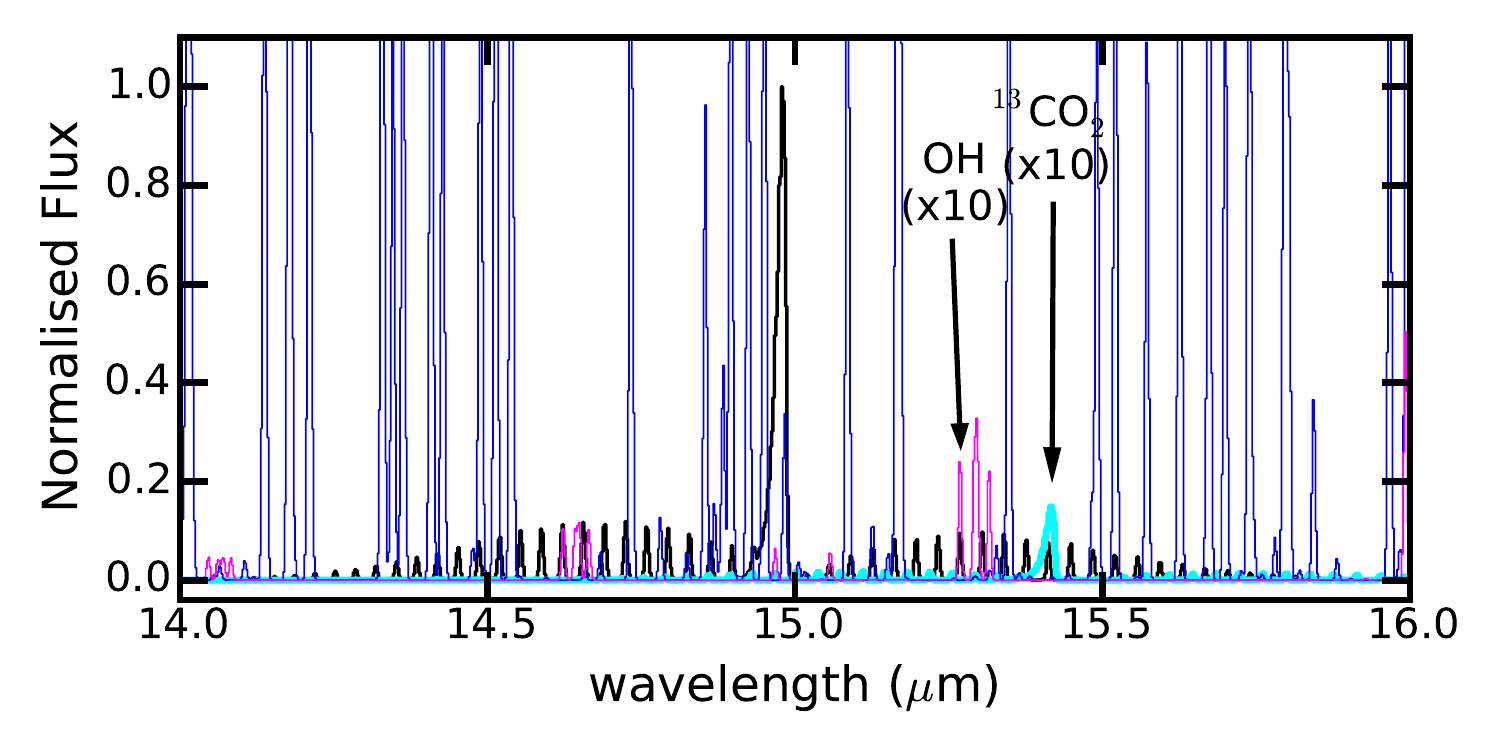}
\caption{\label{fig:CO2_H2Ospectrum} Slab model spectrum comparing {\ce{CO2}} emission (black) and \ce{^{13}CO2} emission (cyan) with the \ce{H2O} emission (blue) and  \ce{OH} emission (magenta) at a resolving power of 2200. Slab models uses the parameters fitted by \cite{Salyk2011} for AS 205 (N) (See. Table~\ref{tab:slabparam}). The large number of strong water lines strongly contaminates the {\ce{CO2}} spectrum. All spectra are normalized to the peak of the {\ce{CO2}} emission. A lot of single water lines are up to four times as strong as the peak of the {\ce{CO2}} 15 $\mu$m feature. The \ce{^{13}CO2} and \ce{OH} fluxes have been multiplied by a factor of 10 for clarity.}
\end{figure}

\begin{figure}
\centering
\includegraphics[width=\hsize]{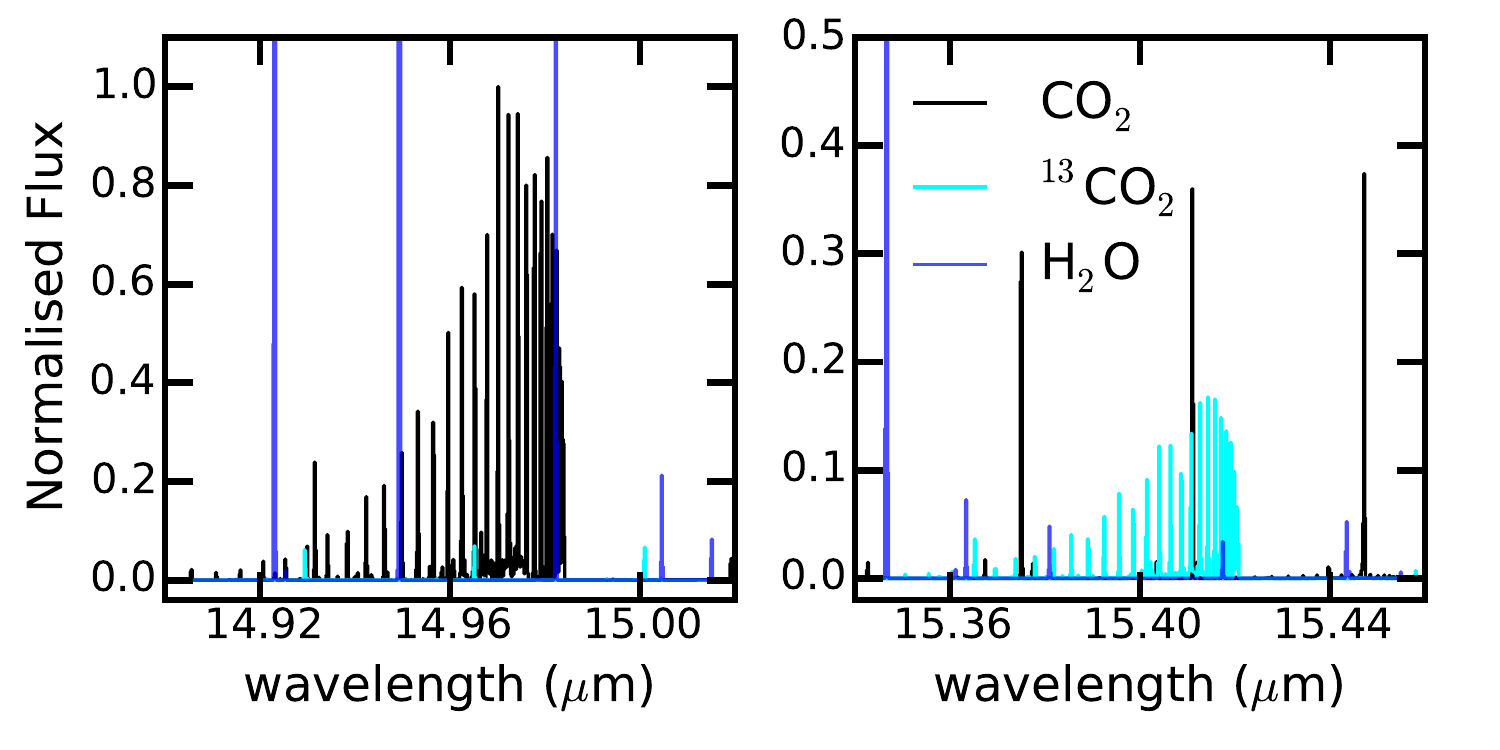}
\caption{\label{fig:CO2_H2Ospectrumhighres} Slab model spectrum comparing {\ce{CO2}} emission (black) and \ce{^{13}CO2} emission (cyan) with the \ce{H2O} emission (blue) at a resolving power of 28000. The left panel shows a zoom of the 15 $\mu$m region, right panel shows a zoom in of the region where the \ce{^{13}CO2} $v_2$ $Q$-branch resides. Slab models uses the parameters fitted by \cite{Salyk2011} for AS 205 (N) (See. Table~\ref{tab:slabparam}). Due to the high resolving power individual $Q$-branch lines can be observed and blends are less likely to happen. A Keplerian linewidth of 20 km s$^{-1}$ has been assumed. The \ce{^{13}CO2} fluxes have been multiplied by a factor of 10 for clarity.}
\end{figure}

One of the major challenges in interpreting IR-spectra of molecules in T-Tauri disks are the ubiquitous water lines. \ce{H2O} has a large dipole moment and thus has strong transitions. As \ce{H2O} chemically favours hot regions \citep{Agundez2008,Walsh2015} there are a lot of rotational lines in the mid infra-red. Fig.~\ref{fig:CO2_H2Ospectrum} shows the \ce{H2O} rotational lines near the {\ce{CO2}} 15 $\mu$m feature. The spectra are simulated with a LTE slab model using the same parameters as fitted by \cite{Salyk2011} for AS 205 (N) as reproduced in Table~\ref{tab:slabparam}. It should be noted that AS 205 (N) is a very water rich disk (in its spectra) explaining the large number of strong lines. Fortunately, there are still some regions in the {\ce{CO2}} spectrum that are not blended with \ce{H2O} or \ce{OH} lines and thus can be used for tracing the {\ce{CO2}} abundance structure independent of a \ce{H2O} emission model. The situation improves at higher resolving power as can be seen in Fig.~\ref{fig:CO2_H2Ospectrumhighres}. The resolving power of 28000 has been chosen to match with the resolving power of the \textit{SPICA} HRS mode. At this point the line widths are dominated by the assumed Keplerian linewidth of 20 km s$^{-1}$. At this resolution the individual $Q$-branch lines are separable and a lot of the line blends that happen at a resolving power of 2200 are no longer an issue. 

\begin{table}
\centering
\caption{\label{tab:slabparam} Parameters for the slab models}
\begin{tabular}{ccc}
\hline
\hline
Molecule & column (cm$^{-2}$) & Temperature (K) \\
\hline
{\ce{CO2}} & $4 \times 10^{15}$ & 300\\
\ce{H2O} & $4 \times 10^{20}$& 300\\
\ce{OH} & $6 \times 10^{16} $& 700\\
\hline
\end{tabular}

\end{table}

\section{\textit{Spitzer}-IRS spectra}
\label{app:spi_spec}
\begin{figure*}
\centering
\includegraphics[width=\hsize]{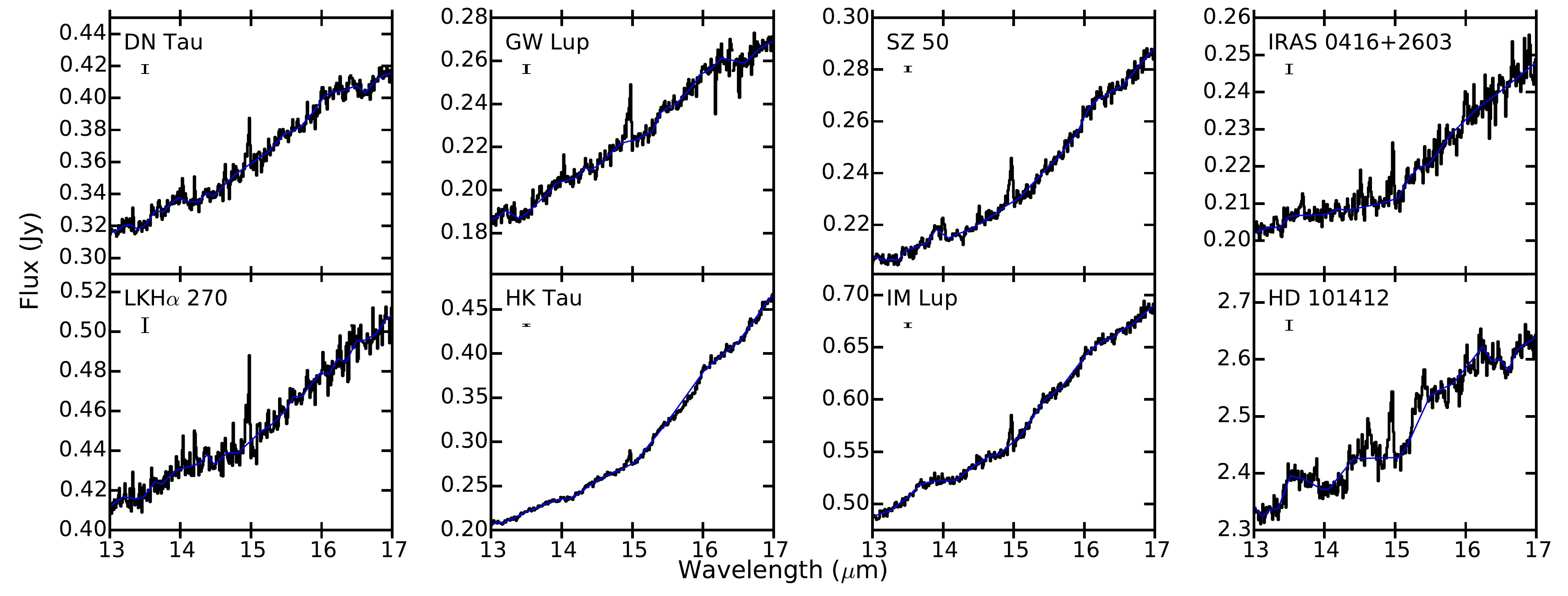}
\caption{\label{fig:spectrum} Observations from \textit{Spitzer-IRS} over 13--17 $\mu$m (black) with continuum as fitted (blue) for the eight selected sources. Typical rms noise on the continuum is shown under the object name. DN Tau, GW Lup and LKH$\alpha$ 270 had a spike in the observed flux at 16.48 $\mu$m due to artefacts at the edge of the observing order. This single data point has been removed from these three spectra.  }
\end{figure*}

Fig.~\ref{fig:spectrum} shows the spectra as observed by \textit{Spitzer}-IRS reduced with the CHIP software \citep{Pontoppidan2010,CHIP2016} in black. The blue lines show the continua that have been fitted to these spectra. The objects have been chosen because their spectra are relatively free of \ce{H2O} emission. Even without \ce{H2O} lines, it is still tricky to determine a good baseline for the continuum as there are a lot of spectral slope changes, even in the narrow wavelength range considered here. This is especially true for HD 101412 where the full spectrum shows a hint of what looks like $R-$ and $P-$branches. If these are features due to line emission, it becomes arbitrary where one puts the actual continuum, thus these features are counted here as part of the continuum. Whether these feature are real or not will not matter a lot for the abundance determination as the \ce{CO2} $Q-$branch is separated from the strong $R-$ and $P-$branch lines.

\begin{figure*}
\centering
\includegraphics[width=\hsize]{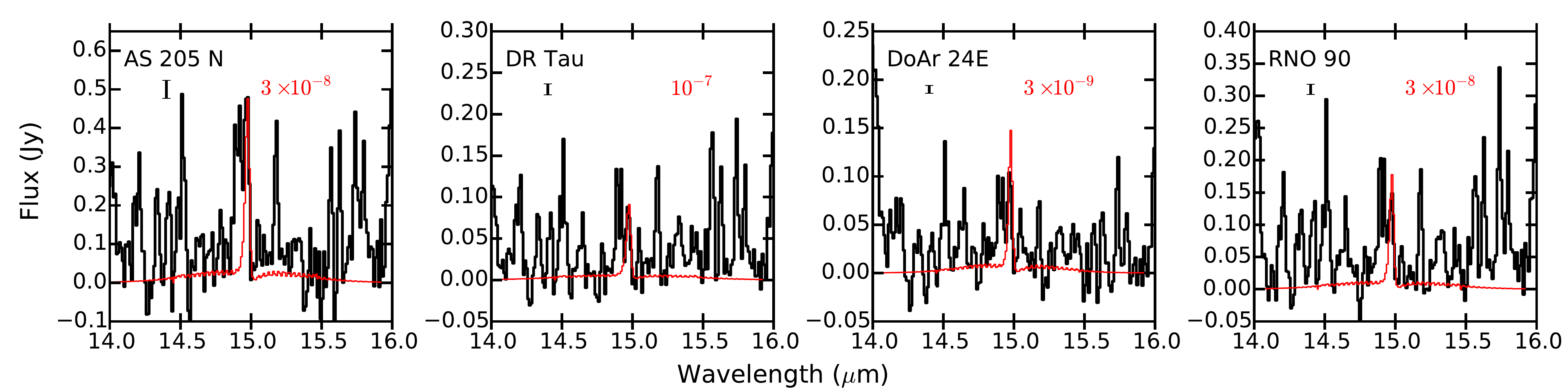}
\caption{\label{fig:Spitzer_H2O} Comparison of continuum subtracted \textit{Spitzer} observations with luminosity and distance corrected model spectra for \ce{CO2}. Object name and median rms noise in the spectra are given in the upper left corner of each panel. All of the sources here have strong \ce{H2O} emission.}
\end{figure*}

Fig.~\ref{fig:Spitzer_H2O} shows a comparison between observed spectra of disk with strong \ce{H2O} emission and \ce{CO2} model spectra. The spectra are corrected for source luminosity and distance as explained in Sec.~\ref{ssc:obs_comp}. Assumed distances and luminosities are given in Table.~\ref{tab:source_water}.

\begin{table}
\centering
\caption{\label{tab:source_water} Stellar parameters}
\begin{tabular}{l c c r}
\hline
\hline
Object & Source & Distance & References \\
& luminosity ($L_\odot$)  & (pc) & \\
\hline
AS 205 N    & 7.3 & 125 & \cite{Andrews2009}\\
DR Tau& 1.0 & 140 & \cite{Rigliaco2015}\\
DoAr24E& 11 & 125 & \cite{Rigliaco2015}\\
RNO 90& 2.7 & 125 & \cite{Rigliaco2015}\\
\hline
\end{tabular}
\end{table}

\end{appendix}

\end{document}